\documentclass[sigconf]{acmart}

\usepackage{booktabs} % For formal tables
\usepackage{graphicx}  %Required
\usepackage{amsmath}
\usepackage{multicol}
\usepackage[labelfont=bf]{caption}
\usepackage{subfig}
\usepackage{float}
\usepackage{amsfonts,algorithmic,algorithm}
\usepackage{booktabs,color}
\usepackage{multirow}
\usepackage{enumitem}
\usepackage{bm}

\newtheorem{theorem}{Problem}
\newtheorem{defn}{Definition}
\newtheorem{theo}{Theorem}

\setcopyright{rightsretained}
\acmDOI{10.475/123_4}

\acmISBN{123-4567-24-567/08/06}

\acmConference[WOODSTOCK'19]{WOODSTOCK}
\acmYear{2019}
\copyrightyear{2016}

\acmArticle{4}
\acmPrice{15.00}

\begin{document}
%\title{Differentially Private Preserving Text Representation Learning}
% \title{Protecting Text Data Privacy through/via Differential Privacy}
%Infer my private information if you can:
%I am not what I write:
%Text Data Anonymization through differential privacy
%Protecting/Anonymizing Text Data through Differential Privacy
%\title{Toward Optimal Privacy-Utility Trade-off: Anonymizing Textual Information by Differential Privacy}
\title[Privacy Preserving Text Representation Learning]{I Am Not What I Write: \\Privacy Preserving Text Representation Learning}

\author{Ghazaleh Beigi, Kai Shu, Ruocheng Guo}
\affiliation{\institution{Computer Science and Engineering, Arizona State University}}
\email{{gbeigi, kaishu, rguo12}@asu.edu}

\author{Suhang Wang}
\affiliation{\institution{College of Information Sciences and Technology, Penn State University}}
\email{swz494@psu.edu}

\author{Huan Liu}
\affiliation{\institution{Computer Science and Engineering, Arizona State University}}
\email{huan.liu@asu.edu}
% The default list of authors is too long for headers.
\renewcommand{\shortauthors}{G. Beigi et al.}

\begin{abstract}

Online users generate tremendous amounts of textual information by participating in different activities, such as writing reviews and sharing tweets. This textual data provides opportunities for researchers and business partners to study and understand individuals. %However, this user-generated textual data can reveal the identity of user and contains individual's private information (e.g., age, location, gender), %leading to privacy leakage.
%compromising the privacy of individuals who provided the textual data.
However, this user-generated textual data not only can reveal the identity of the user but also may contain individual's private information (e.g., age, location, gender). %leading to privacy leakage.
Hence, "you are what you write" as the saying goes. Publishing the textual data thus compromises the privacy of individuals who provided it. The need arises for data publishers to protect people's privacy by anonymizing the data before publishing it. It is challenging to design effective anonymization techniques for textual information which minimizes the chances of re-identification and does not contain users' sensitive information (high privacy) while retaining the semantic meaning of the data for given tasks (high utility). In this paper, we study this problem and propose a novel double privacy preserving text representation learning framework, \textsc{DPText}, %a rigorous solution that searches for the optimal trade-off between privacy and utility. In particular, our approach l
which learns a textual representation that (1) is differentially private, (2) does not contain private information and (3) retains high utility for the given task. Evaluating on two natural language processing tasks, i.e., sentiment analysis and part of speech tagging, we show the effectiveness of this approach in terms of preserving both privacy and utility.
\end{abstract}

%
% The code below should be generated by the tool at
% http://dl.acm.org/ccs.cfm
% Please copy and paste the code instead of the example below.
%

% \ccsdesc[500]{Security and Privacy~Data Anonymization and Sanitization}
% \ccsdesc[300]{Security and privacy~Privacy protections}

%\keywords{Text Representation, Privacy, Utility, Differential Privacy}

\maketitle
% * <huanliu@asu.edu> 2018-11-06T00:23:17.924Z:
% 
% I made some changes to your abstract and introduction on a hard copy, left on your desk. 
% 
% 
% ^ <huanliu@asu.edu> 2018-11-06T00:35:35.672Z.

\section{Introduction}
%researchers infer You are what you write

%The explosive growth of the Web in the last decade has drastically changed the way billions of people all around the globe conduct numerous activities in the Web such as creating online profiles in social media, interacting with other people, and sharing posts in social media platforms. 
\footnote{This is an extended version of the original paper published as a poster paper in the proceedings of the 30th
ACM Conference on Hypertext and Social Media (HyperText'19)~\cite{beigi2019text}}Textual information is one of the most significant portions of data that users generate by participating in different online activities such as leaving online reviews, and posting tweets. On one hand, textual data consists of abundant information about users' behavior, preferences and needs which is critical for understanding individuals by profiling them at unprecedented scales. For example, \textit{data consumers} such as service providers and business partners, use textual data to study customers' behaviors, track users' responses to products, advertise more efficiently, and provide personalized services to users according to their needs. Textual data has been used in many tasks such as sentiment analysis, part-of-speech tagging and information extraction and retrieval~\cite{hovy2015user}. Textual data thus has tremendous usages by various data consumers and have become one of the profitable resources for data publisher~\cite{zhang2016anonymizing,bbc}. %In particular, data consumers request the textual information (e.g., reviews) of all the users who reviewed a specific product. 
%On one hand, textual data consists of abundant information about users' behavior, preferences and needs which is critical for understanding individuals by profiling them at unprecedented scales. It has also been used in many tasks such as sentiment analysis~\cite{zafarani2014social}, part-of-speech tagging~\cite{hovy2015user} and information extraction and retrieval~\cite{zafarani2014social}. %On the other hand, the resultant user-generated textual data is rich in content while it also contains individuals' sensitive and private information, leading to privacy leakage~\cite{narayanan2009anonymizing}.
%On the other hand, the user-generated textual data is rich in content and contains individuals' sensitive and private information, leading to privacy leakage~\cite{narayanan2009anonymizing}. 

On the other hand, %the user-generated textual data is rich in content and recent research has also shown that \textit{"you are what you write"}. This means that 
publishing intact user-generated textual data makes users vulnerable against privacy issues. The reason is that the textual data itself contains sufficient information that allows people in the textual database to be re-identified~\cite{beigi2018privacy,BeigiSigweb,TextAnonymization} and leaks their private attribute information~\cite{mukherjee2010improving,beretta2015interactive,volkova2015inferring}. Thus, "you are what you write" as the saying goes.
%and recent researches have shown that \textit{users are what they write}, meaning that individuals' sensitive and private information could be inferred from the textual data~\cite{beretta2015interactive,jones2007know,volkova2015inferring,mukherjee2010improving}. This leads to leakage of users' privacy. 
 Take the following tweet as an example:%about a user's doctoral appointment as an example:
% This data consists of abundant information about users' behavior which are critical for understanding individuals and profiling them at unprecedented scales. Textual information has been also used in many tasks such as sentiment analysis~\cite{zafarani2014social}, part-of-speech tagging~\cite{hovy2015user} and information extraction and retrieval~\cite{berry2004survey}. On the one hand, the resultant user-generated textual data is rich in content while it also contains individuals' sensitive and private information, leading to privacy leakage~\cite{narayanan2009anonymizing}. Take the following tweet about a user's doctoral appointment as an example:

\textit{Dr.appt Tuesday morning was told I need to lose 30 pounds by X-Mas, have \textbf{high cholesterol}, and \textbf{high blood pressure}. Today starting counting calories \#myfitnesspal and juicing for dinner\footnote{The tweet is real, however, we altered it to preserve the privacy of the user.}}

This user may not be aware that the sensitive medical condition information can be easily inferred from this post-- exposing symptoms of Diabetes. If intact users' textual data is available, a malicious data consumer (or any potential adversary) can easily infer lots of sensitive and private information from text that users' do not explicitly disclose such as vacation plans, medical conditions, age and location~\cite{beretta2015interactive,hovy2015user}. %Users' sensitive and private information that they do not wish to disclose such as vacation plans, medical conditions, age and location can be thus easily inferred from text~\cite{beretta2015interactive,hovy2015user}. %Therefore, publishing textual data could result in inferring sensitive and private information that users do not wish to disclose such as vacation plans, medical conditions, age and location which can be easily inferred from text~\cite{beretta2015interactive,hovy2015user}.
 Another privacy issue arises when a malicious data consumer attempts to re-identify the identity of an individual in the database by investigating whether a targeted user's textual data is in the database or inferring which record is associated with it. %Assume that there is a database which includes tweets from a set of 99 users discussing their plans for Christmas, i.e., staying in town or leaving for vacation. Assume that the data consumer knows 59 users decide to leave tow. After adding a new user's tweets to the database, e.g. Bob, the data consumer figures out that 60 users leave town. The consumer can then easily infer that Bob (the new user) will leave the town for Christmas. 
  %Research has also shown that the textual data itself contains sufficient information that allows people to be re-identified~\cite{jones2007know} and leaks their private attribute information~\cite{volkova2015inferring}.  
  Therefore, publishing complete and intact users' textual data risks exposing their privacy by allowing an adversary to figure out \textit{what} they are.%, i.e.,  in inferring sensitive and private information that users do not wish to disclose such as vacation plans, medical conditions, age and location which can be easily inferred from text~\cite{beretta2015interactive,hovy2015user}. %Thus, releasing complete and intact user textual data risks exposing people's privacy. 

%The leakage of users' information allows any potential adversary, e.g. malicious data consumers, to figure out what the users are. 
These users' privacy concerns, therefore, mandate data publishers to protect privacy by anonymizing the data before sharing it with data consumers. The ultimate goal of an anonymization approach is to preserve user privacy while ensuring the utility of the published data for future tasks and usages. One straightforward technique is to remove ``Personally Identifiable Information'' (a.k.a. PII) such as names and users' IDs. This solution has shown to be insufficient to protect people's privacy. Examples of insufficiencies are the anonymized dataset published for the Netflix prize challenge~\cite{narayanan2008robust} and the AOL search data leak~\cite{barbaro2006face} in which users were re-identified according to their reviews and search queries, respectively. %Research has also shown that the textual data itself contains sufficient information that allows people to be re-identified~\cite{jones2007know} and leaks their private attribute information~\cite{volkova2015inferring}. 
 Various protection techniques for structured data have been developed over the years such as $k$-anonymity and differential privacy. %The ultimate goal of an anonymization approach is to preserve user privacy while ensuring the utility of the published data for future tasks. 
However, traditional privacy preserving techniques are inefficient for user-generated textual data because this data is highly unstructured, noisy and unlike traditional documental content, consists of large numbers of short and informal posts~\cite{fung2010privacy}. Moreover, these works may impose a significant utility loss for protecting textual data as they may not explicitly include utility into the design objective of the privacy protection model. It is thus challenging to design effective anonymization techniques for user-generated textual data which preserves both privacy and utility. 
%  It is challenging to design effective anonymization techniques for user-generated textual data because this data is highly unstructured, noisy and unlike traditional documental content, consists of large numbers of short and informal posts. These challenges make the traditional privacy preserving techniques inefficient for noisy unstructured textual data~\cite{fung2010privacy}.
%This leads to a dilemma of privacy and utility and highlights the need to address the trade-off between them. 

To address the aforementioned challenges, we propose a double privacy preserving text representation learning framework, called \textsc{DPText}. The proposed framework seeks to learn a privacy preserved text representation so that 1) a malicious data consumer (or any potential adversary) cannot infer whether or not a target text representation is in the dataset, %\textcolor{red}{the identity of users in the textual data},
 2) the adversary cannot deduce users' private attribute from the learned representation, and 3) the semantic meaning of the original textual information is still preserved in the learned representation. The learned privacy preserved textual information will be then shared with data consumers.% (e.g. service providers and business partners). %Inspired by the recent success in adversarial learning, Inspired by the idea of adversarial networks.The final text representation is then shared with data consumers (e.g. service providers and business partners).
 
Our double privacy preserving framework protects individuals' privacy against identity re-identification and leakage of private information. Inspired by the recent success in adversarial learning~\cite{goodfellow2014generative}, we build \textsc{DPText} through an integrated process which consists of an auto-encoder, a differential-privacy-based noise adder and two discriminator-learning components (illustrated in Figure~\ref{outline}). We deploy a document auto-encoder to extract latent representation of the original text's content. The noise adder then adds noise to the text representation by adopting a Laplacian mechanism in order to guarantee differential privacy. Although guaranteeing differential privacy minimizes the chances of revealing whether or not a target text representation is in the database, it cannot prevent the adversary from learning user's private information. Moreover, adding too much noise can destroy the semantic meaning of the textual information. To infer the amount of added noise w.r.t. these constraints, we utilize two discriminators that regularize the noise adding process by incorporating necessary constraints. First, we incorporate a \textit{semantic discriminator} to ensure that the semantic meaning of the perturbed text representation is preserved w.r.t. the given task (e.g., classification). Second, we introduce a \textit{private attribute discriminator} to ensure that the perturbed representation does not contain private attributes. %The final text representation is then shared with data consumers (e.g. service providers and business partners).

In essence, we investigate the following challenges: 1) How should textual representation be perturbed to ensure that differential privacy is preserved?, 2) How could we control the amount of the added noise so that the semantic meaning of the text is preserved w.r.t the given task? and 3) How could we handle the amount of the added noise so that the user's private attributes are obscured? Our solution to these challenges results in a novel framework \textsc{DPText}. Our main contributions are summarized as:
\begin{itemize}[leftmargin=*]
	\item We study the problem of text annonymization by learning a differentially private representation that prevents text reconstruction and re-identification by minimizing the chance of attacker to infer whether target text representation is in the database; %re-identification of people who provided the text;% by adding Laplacian noise;%preserves semantic meaning;
	\item We provide a principled way to learn a textual representation %by adopting differential privacy so 
    that does not contain users' private attribute information while retaining the utility for a given task; and
    \item We theoretically show that the learned representation is differentially private which confirms \textsc{DPText} minimizes the re-identification chance. We also conduct experiments on real-world datasets to demonstrate the effectiveness of \textsc{DPText} in two important natural language processing  tasks, i.e., sentiment prediction and part-of-speech (POS) tagging. %, and for protecting several private attributes, i.e., age, gender and location. 
    Our empirical results show that \textsc{DPText} is able to keep the semantic meaning while obscuring private attribute information.%Our results show that \textsc{DPText} does not hurt the semantic meaning of the learned textual representation while obscuring private information.
\end{itemize}
%The rest of this paper is organized as follows:.....
\section{Problem Statement}
We consider an environment with three parties: online users, data publishers, and data consumers. Users generate textual information via various online activities such as posting online information, tweeting and  writing reviews. These information are all collected by data publishers for future usage. A data publisher can be a social media service provider such as Twitter or Facebook or a third-party data company who partners with social media platforms and has access to users' information~\cite{zhang2016anonymizing}. For example, DataSift\footnote{https://datasift.com/} is a third-party company that has access to Twitter's Firehost engine and thus accesses to complete and intact Twitter data including users' tweets. The data publisher can share the collected and anonymize data to data consumers according to users' consent and privacy policies. Data consumers obtain user-generated data by sending requests to data publishers and then use the textual data for understanding individuals at unprecedented scales. %different tasks such as understanding individuals at unprecedented scales, user profiling, information extraction and offering personalized services. 
 Business providers, government agencies and researchers are examples of data consumers. Note that data consumers may not be able to obtain complete and intact user-generated textual data without the support of data publisher.

As we discussed earlier, textual information is rich in content. It can leak users' privacy by allowing users' in the textual database to be re-identified~\cite{TextAnonymization} and leaking their private attribute information~\cite{mukherjee2010improving,beretta2015interactive}. Our focus in this paper is to design an effective text anonymiztion technique for the data publisher to preserve users' privacy by preventing a potential adversary (i.e., malicious data consumer) from breaching privacy of users while maintaining the utility of their textual information for future tasks.% We also assume that the data publisher is trusted by both users and data consumers.

Let $\mathcal{X} = \{x_1, ..., x_N\}$ denotes a set of $N$ documents and $\mathcal{P} = \{p_1, ..., p_T\}$ denotes a set of $T$ private and sensitive attributes. Each document $x_i$ is composed of a sequence of words, i.e., $x_i = \{x_i^1,..., x_i^m\}$. We denote $\mathbf{z}_i \in \mathbb{R}^{d \times 1}$ as the latent representation of the original document $x_i$. We would like to use $x_i$ in the given task $\mathcal{T}$ (e.g., classification). However, we want to preserve users' privacy by preventing a potential adversary from inferring whether a target text representation is in the dataset or which record is associated with it or being able to learn the target users' private attribute information. Thus, in this paper, we study the following problem:
\begin{theorem}
	Given a set of documents $\mathcal{X}$, set of sensitive attributes $\mathcal{P}$, and given task $\mathcal{T}$, learn a function $f$ that can generate and release a manipulated latent representation $\tilde{\mathbf{z}}_i$, for each document $x_i$ so that, 1) the adversary cannot re-identify a targeted text representation and infer whether or not this latent representation is in the database, 2) the adversary cannot infer the targeted user's private attributes $\mathcal{P}$ from the generated representation $\tilde{\mathbf{z}}_i$, and 3) the generated representation $\tilde{\mathbf{z}}_i$ is good for the given task $\mathcal{T}$, i.e., $\tilde{\mathbf{z}}_i = f(x_i, \mathcal{P}, \mathcal{T})$.
\end{theorem}
Note that in our work, the goal is to achieve a protection against possible attacks of malicious data consumers who have access to the released textual information, but not against the system (i.e., text representation learner) which we assume is trusted.\vspace{-5pt}
\section{Background}
\begin{figure*}[ht]\vspace{-10pt}
	\centering
	\includegraphics[width=0.85\linewidth]{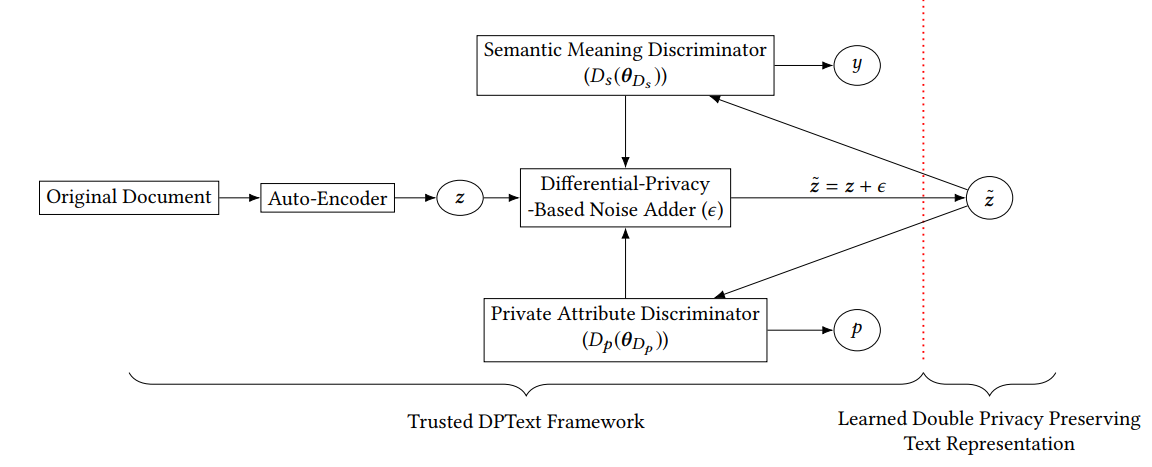}
\vspace{-10pt}
	\caption{The framework of \textsc{DPText} architecture. It consists of four components, a document auto-encoder, a differential-privacy-based noise adder, a semantic meaning discriminator and a private attribute discriminator. We assume that \textsc{DPText} is trusted. Red dashed line shows the privacy barrier and everything to the left of it (i.e., the original data and intermediate results) are kept private. The final learned noisy representation to the right of the privacy barrier is released to the public. This output is a noisy representation which is differentially private, hides private information and has semantic meaning.}%, showing a document training process. The model consists of four components, a document auto-encoder, a differential privacy based noise adder, a semantic meaning discriminator and a private attribute discriminator. The output is a noisy text representation which is differentially private, has semantic meaning and hides private information }}
	\label{outline}\vspace{-5pt}
\end{figure*}
Here, we review the technical preliminaries of differential privacy which is required for the rest of the discussion.
%\subsection{Differential Privacy}
Differential privacy is a powerful technique which protects a user's privacy during statistical query over a database by minimizing the chance of privacy leakage while maximizing the accuracy of queries~\cite{dwork2008differential}. Differential privacy provides a strong privacy guarantee.% and has been leveraged for many privacy preserving application such as graph data~\cite{xiao2014differentially}, textual information~\cite{TextAnonymization} and recommendation systems~\cite{meng2018personalized}. 
%anonymization of graph data~\cite{xiao2014differentially}, textual information~\cite{TextAnonymization}, privacy preserving recommendation systems~\cite{meng2018personalized} and privacy preserving location advertising~\cite{wang2017geographic}.
 The intuition behind differential privacy is that the risk of user's privacy leakage should not increase as a result of participating in a database~\cite{dwork2008differential}. %Differential privacy assumes that data instances are independent from each other.
Differential privacy guarantees that existence of an instance in the database does not pose a threat to its privacy as the statistical information of data would not change significantly in comparison to the case that the instance is absent~\cite{dwork2008differential}. This makes it harder for the adversary to re-identify an instance and infer whether the instance is in the database or not or decide which record is associated with it~\cite{kifer2011no}. We denote an algorithm with privacy property by $\mathcal{A}_p$, which is randomized so that the re-identification of the data on the adversary's side is very difficult. Differential privacy can be formally defined:
% \begin{defn}{\textbf{Differential Privacy}}. Given a query function $f(.)$, a mechanism $K(.)$ with an output range $\mathcal{R}$ satisfies $\epsilon$-differential privacy for all datasets $\mathcal{D}_1$ and $\mathcal{D}_2$ differing in at most one element \textit{iff}:
% 	\begin{equation}\label{dp}
% 		\frac{Pr[K(f(\mathcal{D}_1))=R \in \mathcal{R}]}{Pr[K(f(\mathcal{D}_2))=R \in \mathcal{R}]} \leq e^{\epsilon}
% 	\end{equation}
% \end{defn}
\begin{defn}{\textbf{$\epsilon$ - Differential Privacy}}. An algorithm $\mathcal{A}_p$ is $\epsilon$-differential private if for any subset of outputs $\mathbf{R}$ and for all datasets $\mathcal{D}_1$ and $\mathcal{D}_2$ differing in at most one element:
\begin{equation}\label{dp}
		\frac{\mathbb{P}(\mathcal{A}_p(\mathcal{D}_1) \in \mathbf{R})}{\mathbb{P}(\mathcal{A}_p(\mathcal{D}_2) \in \mathbf{R})} \leq e^{\epsilon}
	\end{equation}
 where $\mathcal{A}_p (\mathcal{D}_1)$ and $\mathcal{A}_p (\mathcal{D}_2)$ are the outputs of the algorithm for input datasets $\mathcal{D}_1$ and $\mathcal{D}_1$, respectively and $\mathbb{P}$ is the randomness of the noise in the algorithm.
\end{defn}
Here $\epsilon$ is called privacy budget and it can be also shown that Eq. \ref{dp} is equivalent to $\lvert \log\big(\frac{P(\mathcal{A}_p(\mathcal{D}_1)=r}{P(\mathcal{A}_p(\mathcal{D}_2)=r}\big)\rvert \leq \epsilon$ for some point $r$ in the output range. Note that larger values of $\epsilon$ (e.g., 10) results in larger privacy loss while smaller values (e.g., $\epsilon \leq 0.1$) indicate the opposite. For example, a small $\epsilon$ means that the output probabilities of $\mathcal{D}_1$ and $\mathcal{D}_2$ at $r$ are very similar to each other which demonstrates more privacy. According to Dwork et al.~\cite{dwork2014algorithmic}, an uncertainty should be introduced in the output of a function (i.e., algorithm) to be able to hide the participation of an individual in the database. This is quantified by sensitivity, which is the amount of the change in the output of function $\mathcal{A}$ made by a single data point in the worst case:
\begin{defn}{\textbf{$L_1$-sensitivity}}. The $L_1$-sensitivity of a vector-valued function $\mathcal{A}$ is the maximum change in the $L_1$ norm of the value of the function $\mathcal{A}$ when one input changes. More formally, the $L_1$-sensitivity $\Delta (\mathcal{A})$ if $\mathcal{A}$ is defined as~\cite{dwork2014algorithmic}:
\begin{align}\label{sensitivity}
\Delta (\mathcal{A}) = \max_{\substack{\mathcal{X},  \mathcal{X}' \\ |\mathcal{X} - \mathcal{X}'| = 1}} \lVert \mathcal{A}(\mathcal{X}) - \mathcal{A}(\mathcal{X}')\rVert_1
% 		\Delta (\mathcal{A}) = \max_i \max_{x_1,..., x_N, x'_i} \lVert \mathcal{A}(x_1, ...,x_{i-1},x_i,x_{i+1}....) \\\nonumber
%         - \mathcal{A}(x_1, ...,x_{i-1},x'_i,x_{i+1},....)\rVert_1%\max_{\substack{x, y \in \mathcal{X} \\ \lVert x - y \rVert_1 = 1}} \lVert \mathcal{A}(x)        - \mathcal{A}(y)\rVert_1
	\end{align}
\end{defn}

% where $\{x_1, ...,x_{i-1},x_i,x_{i+1}....\}$ and $\{x_1, ...,x_{i-1},x'_i,x_{i+1},....\}$ denote two datasets differ in one data point. Next, we discuss the details of our proposed framework \textsc{DPText}.
where $\mathcal{X}$ and $\mathcal{X}'$ are two datasets differ in one entry. Next, we discuss the details of our proposed framework \textsc{DPText}.
\section{The Proposed Framework} Here, we discuss the details of double privacy preserving text representation learning framework. We illustrate the entire model in Figure~\ref{outline}. This framework consists of four major components: 1) an auto-encoder for text representation, 2) differential-privacy-based noise adder, 3) a semantic meaning discriminator, and 4) a private attribute discriminator. The auto-encoder $A$ aims to learn the content representation of a document by minimizing the reconstruction error. Then, the differential-privacy-based noise adder adds a random noise, i.e., Laplacian noise, to the original text representation w.r.t. a given privacy budget to further satisfy the differential privacy guarantee. Since adding noise neither preserves semantic meaning nor necessarily prevents leakage of private attributes, semantic meaning and private attributes discriminators are utilized to infer the amount of the added noise. The semantic meaning discriminator $D_S$ ensures that the added noise does not destroy the semantic meaning w.r.t. a given task. The private attribute discriminator $D_P$ also guides the amount of added noise by ensuring that the manipulated representation does not include users' private information. Note that we assume that the framework is trusted and therefore everything to the left of the privacy barrier (the red dashed line in Figure~\ref{outline}) including the original textual information and intermediate results, are kept private. The final learned representation which is to the right of the privacy barrier is released to the public. The final output 1) is differentially private, 2) obscures private attribute information, and 3) preserves semantic meaning.%We will show later that the resultant representation satisfies differential privacy.%Next, we will discuss each component in details.
%\subsection{Encoder for Text Representation}
\subsection{Extracting Textual Representation}\label{encoder}
Here, we demonstrate how to extract the content representation for a given document. Let $x = \{x^1,..., x^m\}$ be a textual document with $m$ words. Auto-encoder has been widely utilized for text generation and has shown to be effective recently~\cite{bowman2015generating,cho2014learning}. 
%RNN based auto-encoder is known for its ability in summarizing and learning semantic of unstructured noisy short texts~\cite{cho2014learning,shang2015neural}.
 We therefore use an auto-encoder $A$ to extract content representation $\mathbf{z}$ from document $x$. Let $E_A : \mathcal{X} \rightarrow \mathcal{Z}$ be an encoder that can infer the content representation $\mathbf{z}$ for a given document $x$, and $D_A : \mathcal{Z} \rightarrow \mathcal{X}$ be a decoder that reconstruct the document from its learned representation.

Recurrent neural networks (RNNs) have been shown to be effective for summarizing and learning semantic of unstructured noisy short texts~\cite{cho2014learning,shang2015neural}. %modeling textual information since it can capture semantic meaning of texts. 
 In this work, we apply RNN as the encoder to learn the latent representation of texts. RNN can learn a probability distribution over a sequence by being trained to predict the next symbol in a sequence. The RNN consists of a hidden state $S$ and an optional output which operates on a word sequence $x = \{x^1,..., x^m\}$. At each time step $t$, the hidden state $s_t$ of RNN is updated by,
\begin{equation}
	s_t = f_{enc}(s_{t-1}, x^t)
\end{equation}

After reading the end of the given document, we use the last hidden state of the RNN as the representation vector $\mathbf{z} \in \mathbb{R}^{d \times 1}$ of the document $x$. We employ the gated recurrent unit (GRU) as the cell type to build the RNN, which is designed in a manner to have a more persisted memory~\cite{cho2014learning}. Let $\theta_e$ denotes the parameters for the encoder $E_A$. Then we will have:
\begin{equation}
	\mathbf{z} = E_A(x, \theta_e)\label{latentrep}
\end{equation}

Decoder $\hat{x} = D_A(\mathbf{z}, \theta_d)$ takes $\mathbf{z}$ as the input to start the generation process and $\theta_d$ denotes the parameters for the decoder $D_A$. We use another RNN to build the decoder $D_A$ to generate the output word sequence $\hat{x} = \{\hat{x}^1,.., \hat{x}^m\}$. At each time step $t$, the hidden state of the decoder is computed as:
\begin{equation}
	s_t = f_{dec}(s_{t-1}, \hat{x}^t)
\end{equation}
where $s_0=\mathbf{z}$. The two components of the proposed auto-encoder are jointly trained to minimize the negative conditional log-likelihood for all documents. We use the trained auto-encoder $E_A$ to obtain the content representation $\mathbf{z} \in \mathbb{R}^{d \times 1}$ according to Eq. \ref{latentrep} where $d$ is the size of textual representation.
\subsection{Preventing Text Re-identification and Reconstruction by Adding Noise}
Textual information is rich in content and publishing this data without proper anonymization lead to privacy breach and revealing the identity of an individual. This can let the adversary infer if a targeted user's latent textual representation is in the database or which record is associated with it. Moreover, publishing a document's latent representation could result in leakage of the original text. In fact, recent advancement in adversarial machine learning shows that it is possible to recover the input textual information from its latent representation~\cite{hitaj2017deep}. In this case, if an adversary has preliminary knowledge of the training model, they can readily reverse engineer the input, for example, by a GAN attack algorithm~\cite{hitaj2017deep}. It is thus essential to protect the textual information before publishing it.

Differential privacy is a powerful technique for preserving privacy of users' data included in a database and provides a privacy guarantee. Our method is inspired by Chaudhuri et al.~\cite{chaudhuri2011differentially}, where the differential privacy is achieved through adding a random noise, i.e., Laplacian noise, to the output of an algorithm $\mathcal{A}$. This mechanism is known as \textit{output perturbation} and it has been proved that under certain conditions this output perturbation mechanism will guarantee differential privacy~\cite{chaudhuri2011differentially}. 

%Output perturbation is a very popular differential privacy based approach to protect the raw data. More specifically, 
The main idea of the output perturbation mechanism is to add noise to the output of an algorithm to preserve its privacy. In our problem, the output is the original document latent representation $\mathbf{z}$. The benefit of adding noise to this latent representation is two fold. First, it minimizes the chance of the re-identification of learned text representation by preventing the adversary to infer whether or not a target representation is in the database, and second, it makes it difficult for the adversary to recover the raw textual data. The goal here is thus to add noise to the output such that the differential privacy condition is satisfied. Laplacian mechanism is a popular way to add noise to preserve differential privacy. %In this paper, we adopt Laplacian mechanism. However, it is very straightforward to adopt Guassian mechanism under the proposed framework. 
In particular, with Laplacian mechanism, we perturb the output $\mathbf{z}$ by adding Laplacian noise to it as follows:
% \begin{equation}
% \tilde{z} = z+s,~~~~~s \sim \mathcal{N}(\mu, diag(\sigma))
% \end{equation}
\begin{equation}
	\tilde{\mathbf{z}}(i) = \mathbf{z}(i)+\mathbf{s}(i),~~~\mathbf{s}(i) \sim Lap(b), ~~ b = \frac{\Delta}{\epsilon},~~i=1,..,d\label{Laplacian}
%    \tilde{\mathbf{z}}(i) = \mathbf{z}(i)+\mathbf{s}(i),~~i=1,..,d
\end{equation}
where $\epsilon$ is the privacy budget, $\Delta$ is the $L_1$-sensitivity of the latent representation $\mathbf{z}$, $d$ the dimension of $\mathbf{z}$, $\mathbf{s}$ the noise vector, $\mathbf{s}(i)$ and $\mathbf{z}(i)$ are the $i$-th element for vectors $\mathbf{s}$ and $\mathbf{z}$, respectively. $\Delta = 2d$ (see details in Section \ref{proof_sec}). Note that each element of the noise vector is drawn from Laplacian distribution. %In Section \ref{proof_sec}, we will show that $\Delta = 2d$.

\subsection{Preserving Semantic Meaning}
Perturbing the latent representation of the given text by adding noise to it (Eq.~\ref{Laplacian}) prevents the adversary from re-constructing the text from its latent representation and guarantees differential privacy. However, this approach may destroy the semantic meaning of the text data. Semantic meaning is task-dependant, e.g., classification is one of the common tasks. In the case of sentiment analysis, sentiment is of semantic meaning in the given text and sentiment prediction is a classification task. In order to preserve the semantic meaning of the textual representation, we need to add an optimal amount of noise to the text latent representation which does not destroy the semantic meaning of the text data while ensuring data privacy. We approach this challenge by \textit{learning} the amount of the added noise with the privacy budget $\epsilon$ in terms of training a classifier:
\begin{equation}
	\hat{y} = softmax(\tilde{\mathbf{z}};\theta_{D_S})
\end{equation}
where $\theta_{D_S}$ are the weights associated with the softmax function and $\hat{y}$ represents the inferred label for the classification. 

To preserve the semantic meaning of the text representation, we seek a noisy latent representation which retains high utility and accordingly contains enough information for a downstream task, e.g., classification. %In essence, we want to learn the amount of the added noise so that the resultant latent representation be good for the given classification task. 
We define a \textit{semantic discriminator} $D_S$ that aims to assign a correct class label to the perturbed representation, whose loss function is minimized as follows,
\begin{align}
	&\min_{\theta_{D_S}, \epsilon} \mathcal{L}(\hat{y}, y) = \min_{\theta_{D_S}, \epsilon} \sum_{i=1}^{C} -y(i)\log \hat{y}(i)%\quad s.t.~~~~\epsilon \le c_1
    \label{utility}
\end{align}
where $C$ is the number of classes, and $\mathcal{L}$ denotes the cross entropy loss function. The one-hot encoding of the ground truth label for the classification task is also denoted by $y$ and $y(i)$ represents the $i$-th element of $y$, i.e., the ground truth label for $i$-th class.

To learn the value of the privacy budget $\epsilon$, we employ the commonly used reparameterization trick~\cite{kingma2013auto}. Instead of directly sampling noise $\mathbf{s}(i)$ from Laplacian distribution (i.e., Eq. \ref{Laplacian}), this trick first samples a value $r$ from a uniform distribution, i.e. $r \sim [0,1]$, and then rewrites the amount of added noise $\mathbf{s}(i)$ as follows:
\begin{equation}
	\mathbf{s}(i) = -\frac{\Delta }{\epsilon} \times sgn(r) \ln(1-2|r|), \quad i = 1,2,..,d\label{noise}
\end{equation}
This is equivalent to sampling noise $s$ from $Lap(\frac{\Delta}{\epsilon})$.
% \begin{equation}
% \tilde{z} = z+\mu + \epsilon \odot \sigma,~~~~~\epsilon \sim \mathcal{N}(0, 1)\label{noise}
% \end{equation}
% The advantage of doing this is that the parameters $\mu$ and $\sigma$ are now explicitly involved in the representation of added noise $s$ which makes it possible to use back-propagation to find the optimal value of the parameters $\mu$ and $\sigma$.
The advantage of doing so is that the parameter $\epsilon$ is now explicitly involved in the representation of the added noise, $\mathbf{s}$, which makes it possible to use back-propagation to find the optimal value of $\epsilon$. Large privacy budget $\epsilon$ could result in large privacy bounds. Hence, we add a constraint, $\epsilon<c_1$ where $c_1$ is a predefined constraint. 

Another challenge here is that, $\hat{y}$ is inferred from $\tilde{\mathbf{z}}$ after introducing noise to the original latent representation $\mathbf{z}$. The noise is also sampled from the Laplacian distribution which results in large variance in the training process. To solve this issue and make the model more robust, we sample $K$ copies of noise for each given document. In other words, we can rewrite Eq.~\ref{utility} as follows:
\begin{align}\label{utility_revised}
	%\min_{\theta_e, W, \epsilon}  \mathcal{L}_{D_S} \\\ \nonumber
	&\min_{\theta_{D_S}, \epsilon} \mathcal{L}_{D_S}(\hat{y}, y) = \min_{\theta_{D_S}, \epsilon} \frac{1}{K}\sum_{k = 1}^{K}\mathcal{L}(\hat{y}^k, y) = \\\nonumber
    &\min_{\theta_{D_S}, \epsilon} \frac{1}{K}\sum_{k = 1}^{K} \sum_{i=1}^{C} -y(i)\log \hat{y}^k(i)\quad s.t. \quad \epsilon \le c_1
\end{align}
where the goal is to minimize loss function $ \mathcal{L}_{D_S} $ w.r.t. the parameters $\{\theta_{D_S}, \epsilon\}$, and $\hat{y}^k = softmax(\tilde{\mathbf{z}}^k;\theta_{D_S})$. Note that $\tilde{\mathbf{z}}^k = \mathbf{z}+\mathbf{s}^k$ in which $\mathbf{s}^k$ is the $k$-th sample of the noise calculated with Eq.~\ref{noise}.

\subsection{Protecting Private Information}
We discuss how adding noise to the latent representation of the text can prevent adversary from learning the input textual information and guarantee differential privacy. Another important aspect of learning privacy preserving text representation is to ensure that sensitive and private information of the users such as age, gender, and location is not captured in the latent representation. 

%To address this issue, we propose to have a textual representation which not only is accurate w.r.t. the given task but also has maximum privacy against the strongest attacker. 
An adversary cannot design a private attribute inference attack better than what it has already anticipated. In this spirit, we leverage the idea of adversarial learning. In particular, we seek to train a \textit{private attribute discriminator} $D_P$ that can accurately identify the private information from the given representation, while learning a representation that can fool the discriminator and minimize leakage of private attribute w.r.t. the determined adversary, which results in a representation that does not contain sensitive information. Assume that there are $T$ private attributes (e.g., age, gender, location). Let $p_t$ represents the ground truth (i.e., correct label) for the $t$-th sensitive attribute and $\theta_{D_P^t}$ demonstrates the parameters of discriminator model $D_P$ for the $t$-th sensitive attribute. The adversarial learning can be formally written as:\vspace{-10pt}
\begin{align}\label{privacy}
	\min_{\{\theta_{D_P^t}\}^T_{t=1}} \max_{\epsilon} \mathcal{L}_{D_P} = \min_{\{\theta_{D_P^t}\}^T_{t=1}} \max_{ \epsilon} \frac{1}{K.T}\sum_{t=1}^{T}\sum_{k = 1}^{K}\mathcal{L}_{D_P^t}(\hat{p}^k_t, p_t),~~~~s.t. ~~\epsilon \le c_1
\end{align}%\vspace{-10pt}
where $\mathcal{L}_{D_P^t}$ denotes the cross entropy loss function and $\hat{p}^k_t = softmax(\tilde{\mathbf{z}}^k, \theta_{D_P^t})$ is the predicted $t$-th sensitive attribute using the $k$-th sample. The outer minimization finds the strongest private attribute inference attack and the inner maximization seeks to fool the discriminator by obscuring private information.
\subsection{DPText - Learning the Text Representation}
In the previous sections, we discuss how we can (1) add noise to prevent the adversary from reconstructing the original text from the latent representation and minimize the chance of privacy breach by satisfying differential privacy (Eq.~\ref{Laplacian}), (2) control the amount of the added noise to preserve the semantic meaning of the textual information for a given task (Eq.~\ref{utility_revised}), and (3) control the amount of the added noise so that user's private information is masked  (Eq.~\ref{privacy}). Inspired by the idea of adversarial learning, we achieve all three by modeling the objective function as a minmax game among the two introduced discriminators as follows:
\begin{align}
	&\min_{\theta_{D_S}, \epsilon} \max_{\{\theta_{D_P^t}\}^T_{t=1}} \mathcal{L}_{D_S} - \alpha \mathcal{L}_{D_P} = \\ \nonumber\label{Obj_one}
	&\min_{\theta_{D_S}, \epsilon} \max_{\{\theta_{D_P^t}\}^T_{t=1}}\frac{1}{K}\sum_{k = 1}^{K} \bigg[\mathcal{L}(\hat{y}^k, y) - \alpha \frac{1}{T}\sum_{t = 1}^{T}\mathcal{L}_{D_P^t}(\hat{p}^k_t, p_t)\bigg],~~~~~ s.t. ~~~ \epsilon \le c_1
\end{align}
where $\alpha$ controls the contribution of the private attribute discriminator in the learning process. This objective function seeks to minimize privacy leakage w.r.t. the attack, minimize loss in the semantic meaning of the textual representation, and protect private information. %This results in designing the private attribute information preserving mechanism which is also semantic meaning maximizing. 
%Assuming we have $N$ documents, 
With $N$ documents, Eq.~12 %\ref{Obj_one} 
is written as follows:
\begin{align}
	% &\min_{\theta_e, W, \epsilon} \max_{\{\theta_{D_t}\}^T_{t=1}} \frac{1}{N}\sum_{n = 1}^{N}\big[\mathcal{L}_{D_S} - \mathcal{L}_{D_P}\big] = \\ \nonumber 
	& \min_{\theta_{D_S}, \epsilon} \max_{\{\theta_{D_P^t}\}^T_{t=1}}\frac{1}{N}\sum_{n = 1}^{N}\bigg[\frac{1}{K}\sum_{k = 1}^{K} \bigg[\mathcal{L}(\hat{y}^k_n, y_n) - \\ \nonumber
	&~~~~~~~~~~~~~~~~~~~~~~\alpha \frac{1}{T}\sum_{t = 1}^{T}\mathcal{L}_{D_P^t}(\hat{p}^k_{n,t}, p_{n,t})\bigg] \bigg]
	+\lambda \Omega(\theta) 
	\quad \quad s.t. \quad \quad \epsilon \le c_1
%	\label{Obj_all}
\end{align}
where $\theta= \{\theta_{D_S}, \epsilon, \{\theta_{D_P^t}\}^T_{t=1}\}$ is the set of all parameters to be learned, $\Omega(\theta)$ is the regularizer for the parameters such as Frobenius norm and $\lambda$ is a scalar to control the amount of contribution of the regularization $\Omega(\theta)$.

The aim of this objective function is to perturb the original text representation by adding a proper amount of noise to it in order to prevent an adversary from inferring existence of the target textual representation in the database, reconstructing the user's original text and learning user's sensitive information from the latent representation, while preserving the semantic meaning of the perturbed representation for a given specific task. We stress that the resultant text representation satisfies $\tilde{\epsilon}$-differential privacy, where $\tilde{\epsilon} \leq c_1$ is the optimal learned privacy budget. This is further discussed in Section.~\ref{proof_sec}. %Note that the final perturbed latent text representation is released at the end.
\begin{algorithm}[t]
	\caption{The Learning Process of \textsc{DPText} model}\label{learning}
	\begin{algorithmic}[1]
		\REQUIRE~~ Training data $\mathcal{X}$, $\theta_{D_S}$, $\epsilon$, $\{\theta_{D_P^t}\}^T_{t=1}$, batch size $b$, $c_1$ and $\alpha$.
        \ENSURE~~ The privacy preserving learned text representation $\tilde{\mathbf{z}}$
% 		\ENSURE~~  The optimal privacy budget value $\tilde{\epsilon}$, the sentiment discriminator generator $D_S$, and the private discriminator $D_P$.
		\STATE Pre-train  the document auto-encoder $E_A$ to obtain the text representations according to Eq. \ref{latentrep} as $\mathbf{z}=E_A(x,\theta_e)$ \label{alg:vae}
		\REPEAT \label{alg:repeat}
		\STATE Sample a mini-batch of $b$ samples $\{x^i\}_{i=1}^b$ from $\mathcal{X}$
		\STATE Add noise $\mathbf{s}$ to initial document representation $\mathbf{z}_i$ and get the new document representation $\tilde{\mathbf{z}}_i$, $i = 1,2,...,b$ via Eq.\ref{noise}\label{alg:professor}
		\STATE Train semantic discriminator $D_S$ by gradient descent(Eq.\ref{utility_revised})\label{alg:dt}
		\STATE Train private attribute discriminator $D_P$ via Eq.\ref{privacy}.\label{alg:lossg}
		\UNTIL{\text{Convergence}}  \label{alg:until}
%         \STATE For each document latent representation $\mathbf{z}_n$, $n=1,...,N$, calculate $\tilde{\mathbf{z}}_n = \mathbf{z}_n + $
	\end{algorithmic}\label{alg:opt}
\end{algorithm}
\subsection{Optimization Algorithm}
The optimization process is illustrated in Algorithm~\ref{alg:opt}. First, we compute the latent representation of all documents $\mathcal{Z} = \{\mathbf{z}_i, ..., \mathbf{z}_N\}$ in Line~\ref{alg:vae}. We then sample a mini-batch of $b$ samples from the training data and add noise to initial to initial text representation. Next, we train the semantic discriminator $D_S$ in Line~\ref{alg:dt} and private attribute discriminator %using Eq.~\ref{privacy}
in Line~\ref{alg:lossg}. Recall that we have a constraint on the variable $\epsilon$, i.e., $\epsilon<c_1$. To satisfy this constraint, we use the idea of the projected gradient descent~\cite{boyd2004convex} wherein the gradient descent is performed one step, i.e. $\epsilon-\gamma\times\epsilon$ where $\gamma$ is the learning rate. Then, the parameter $\epsilon$ is projected back to the constraint. This means that if $\epsilon >c_1$, then we set $\epsilon = c_1$, otherwise, keep the value of $\epsilon$. The final noisy representation $\tilde{\mathbf{z}}$ can be then calculated for each given document according to the value of optimal learned privacy budget $\tilde{\epsilon} \leq c_1$ using Eq.  \ref{Laplacian}. %Next, we will prove that for each document $n$, the noisy latent representation $\tilde{\mathbf{z}}_n$ satisfies $\tilde{\epsilon}$-differential privacy.
 Note that any model can be used for semantic and private attribute discriminators.
%Note that exploring the best POS tagger is not the focus of our work and it can be easily replaced by different models designed for this task.
%the models used for POS tagger and private attribute discriminator can be easily replaced by different models designed for these tasks.
\section{Theoretical Analysis}\label{proof_sec}
Here, we show that the learned text representation using \textsc{DPText} is $\tilde{\epsilon}$-differential privacy where $\tilde{\epsilon} \leq c_1$ is the learned optimal privacy budget. In particular, we prove the privacy guarantee for the final noisy latent representation $\mathbf{\tilde{z}}$ for each given document. The theoretical findings confirm the fact that \textsc{DPText} minimizes the chance of revealing existence of textual representations in the database.\vspace{-5pt}
%In this section, we prove the privacy guarantee for the final noisy latent representation $\mathbf{\tilde{z}}$ for each given document.
\begin{theo}
Let $\tilde{\epsilon} \leq c_1$ be the optimal value learned for the privacy budget variable $\epsilon$ w.r.t the semantic meaning and private attribute discriminators. Let $\mathbf{z}_i$ be the original latent representation for document $\mathbf{x}_i$, $i = 1,..., N$ inferred using Eq.~\ref{latentrep} and. Moreover, let $\Delta$ denotes the $L_1$-sensitivity of the textual latent representation extractor function discussed in Section.~\ref{encoder}. If each element $\mathbf{s}_i(l)$, $l = 1,..., d$ in noise vector $\mathbf{s}_i$ is selected randomly from $Lap(\frac{\Delta }{\tilde{\epsilon}})$ ($\Delta = 2d$), the final noisy latent representation $\mathbf{\tilde{z}}_i = \mathbf{z}_i + \mathbf{s}_i$ satisfies $\tilde{\epsilon}$-differential privacy.
\end{theo}\vspace{-5pt}
\begin{proof}
First we bound the change of $\mathbf{z}$ when one data point in the database changes. This gives the $L_1$-sensitivity of the textual latent representation extractor function discussed in Section. \ref{encoder}.

Recall the way $\mathbf{z}$ is calculated using Eq.~\ref{latentrep}. Function $tanh$ is used in GRU to build the RNN which is used in Section. \ref{encoder} to find the latent representation of a given document. The output of $tanh$ function is within range $[-1,1]$. This indicates that value of each element $\mathbf{z}(l)$, $l=1,...,d$ in the latent representation vector $\mathbf{z}$ is within range $[-1,1]$. If one data point changes (i.e., removed from the database), the maximum change in value of each element $\mathbf{z}(l)$ is 2. Since the dimension of $\mathbf{z}$ is $d$, the maximum change in the $L_1$ norm of $\mathbf{z}$ happens when all of its elements, $\mathbf{z}(l)$, have the maximum change. According to Definition. \ref{sensitivity}, the $L_1$-sensitivity of $\mathbf{z}$ is $\Delta = 2\times d$. 

Now, assume that $\tilde{\epsilon} \leq c_1$ is the optimal value for the learned privacy budget. Then each element in $\mathbf{s}$ (i.e., $\mathbf{s}(l), ~~ l = 1,2,...d$) is distributed as $Lap(\frac{\Delta}{\tilde{\epsilon}})$ based on Eq. \ref{Laplacian} which is equal to randomly picking each $\mathbf{s}(l)$ from the $Lap(\frac{\Delta}{\tilde{\epsilon}})$ distribution, whose probability density function is $Pr(\mathbf{s}(l)) = \frac{\tilde{\epsilon}}{2\Delta } e ^{- \frac{\tilde{\epsilon} \lvert \mathbf{s}(l)\rvert}{\Delta }}$.

Let $\mathcal{D}_1$ and $\mathcal{D}_2$ be any two datasets only differ in the value of one record. Without loss of generality we assume that the representation of the last document is changed from $\mathbf{z}_n$ to $\mathbf{z}'_n$. Since the $L_1$-sensitivity of $\mathbf{z}$ is $\Delta = 2d$, then $\lVert \mathbf{z}_n -  \mathbf{z}'_n \rVert_1 \leq \Delta$. Then we have:
\begin{align}\label{proof_eq}
&\frac{Pr[\mathbf{z}_n + \mathbf{s}_n = r|\mathcal{D}_1]}{Pr[\mathbf{z}'_n+ \mathbf{s}'_n = r|\mathcal{D}_2]} =\frac{\prod_{l \in \{ 1,2,...,d\}} Pr(r - \mathbf{z}_n(l))}{\prod_{l \in \{ 1,2,...,d\}} Pr(r - \mathbf{z}'_n(l))} \\ \nonumber
& = \frac{\prod_{l \in \{ 1,2,...,d\}} Pr(\mathbf{s}_n(l))}{\prod_{l \in \{ 1,2,...,d\}} Pr(\mathbf{s}'_n(l))}  = e ^{- \frac{\tilde{\epsilon} \sum_l \lvert \mathbf{s}_n(l)\rvert}{\Delta }} / e ^{- \frac{\tilde{\epsilon} \sum_l \lvert \mathbf{s}'_n(l)\rvert}{\Delta }} \\ \nonumber
&= e ^{ \frac{\tilde{\epsilon} \sum_l (\lvert \mathbf{s}'_n(l)\rvert - \lvert \mathbf{s}_n(l)\rvert)}{\Delta }}  \leq e ^{\frac{\tilde{\epsilon} \sum_l \lvert \mathbf{s}'_n(l) - \mathbf{s}_n(l)\rvert}{\Delta}} = e ^{\frac{\tilde{\epsilon}  \lVert \mathbf{s}'_n - \mathbf{s}_n\rVert_1}{\Delta }}
\end{align}
 where $\mathbf{s}_n$ and $\mathbf{s}'_n$ are the corresponding noise vectors with respect to the learned $\tilde{\epsilon}$ when the input are $\mathcal{D}_1$ and $\mathcal{D}_2$, respectively. The first inequality also follows from the triangle inequality, i.e. $\lvert a \rvert - \lvert b \rvert \leq \lvert a - b \rvert$. The last equality follows from the definition of $L_1$-norm.
 
 Since we have $\mathbf{s}_n = r - \mathbf{z}_n$ and  $\mathbf{s}'_n = r - \mathbf{z}'_n$, we can write:
 \begin{equation}
 \lVert \mathbf{s}'_n - \mathbf{s}_n\rVert_1 = \lVert (r - \mathbf{z}'_n )-( r - \mathbf{z}_n)\rVert_1 = \lVert \mathbf{z}'_n - \mathbf{z}_n\rVert_1 \leq \Delta
 \end{equation}
 This follows from the definition of $L_1$-sensitivity. We rewrite Eq.~\ref{proof_eq}:
 \begin{align}
 \frac{Pr[\mathbf{z}_n + \mathbf{s}_n = r|\mathcal{D}_1]}{Pr[\mathbf{z}'_n+ \mathbf{s}'_n = r|\mathcal{D}_2]} \leq e ^{\frac{\tilde{\epsilon}  \lVert \mathbf{s}'_n - \mathbf{s}_n\rVert_1}{\Delta }} \leq e ^{\frac{\tilde{\epsilon}   \Delta }{\Delta }} = e ^{\tilde{\epsilon}}
 \end{align}
 
So, the theorem follows and the final noisy latent representation is $\tilde{\epsilon}$-differentially private.
\end{proof}
\section{Experiments}
In this section, we conduct experiments on real-world data to demonstrate the effectiveness of \textsc{DPText} in terms of preserving both privacy of users and utility of the resultant representation for a given task. Specifically, we aim to answer the following questions:

\begin{itemize}[leftmargin=*]
	\item \textbf{Q1} - \textit{Utility}: Does the learned text representation preserve the semantic meaning of the original text for a given task?
	\item \textbf{Q2} - \textit{Privacy}: Does the learned text representation obscure users' private information?
    \item \textbf{Q3} - \textit{Utility-Privacy Relation}: Does the improvement in privacy of learned text representation result in sacrificing the utility?
%  \item \textbf{Q3} - \textit{Utility-Privacy Relation}: How successful is \textsc{DPText} in handling privacy-utility trade-off?
\end{itemize}
To answer the first question (\textbf{Q1}), we report experimental results for \textsc{DPText} w.r.t. two well known text-related tasks, i.e., sentiment analysis and part-of-speech (POS) tagging. Sentiment analysis and POS tagging have many applications in Web and user-behavioral modeling~\cite{hovy2015tagging,jorgensen2016learning}. %These information can also raise privacy issues for people whose texts are analyzed and used to train models. 
Recent research showed how linguistic features such as sentiment are highly correlated with users demographic information~\cite{hovy2015user,potthast2017overview}. Another group of research shows the effectiveness of POS tags in predicting users' age and gender information~\cite{mukherjee2010improving}. This makes users vulnerable against inference of their private information. Therefore, to answer the second question (\textbf{Q2}), we consider different private information, i.e., age, location, and gender, and report results for private attribute prediction task. To answer the third question (\textbf{Q3}), we investigate the utility loss against privacy improvement of the learned text representation. %Next, we discuss each task and the experimental settings.
\subsection{Task 1: Sentiment Analysis}
Sentiment analysis is one of the important language processing applications. Next, we describe the used dataset and model.
\subsubsection{Data}
We use a dataset from TrustPilot~\footnote{http://trustpilot.com} from Hovy et al.~\cite{hovy2015user}. On their website, users can write reviews and leave a one to five star rating. Users can also provide some demographic information.% This website is available for 24 countries, and in 13 different languages. 
 In the collected dataset, each review is associated with three attributes, gender (male/female), age, and location (Denmark, France, United Kingdom, and United States). %HL? To ensure that location is not predictable based on the script, 
 We follow the same approach as in~\cite{li2018towards} and discard all non-English reviews based on \textsc{LANGID.PY}\footnote{https://github.com/saffsd/langid.py}~\cite{lui2012langid}, and only keep reviews classified as English with a confidence greater than 0.9. We follow the setting of~\cite{hovy2015tagging} and categorize age attribute into three groups, over-45, under-35, and between 35 and 45. %Moreover, we discard reviews that do not have information for all three attributes. 
We follow the setting of~\cite{lui2012langid} and subsample 10k reviews for each location to balance the five locations.%, which were highly skewed in the original dataset. 
We consider each review's rating score as the target sentiment class. %The statistics of the preprocessed data are shown in Table~\ref{data}
% \begin{table*}
% \caption{\textbf{Statistics of the preprocessed TrustPilot data. The numbers in the table show the number of reviews. The numbers in the parenthesis demonstrates the average review rating for the corresponding group }}\label{data}
%   \begin{tabular}{lSSSSS}
%     \toprule
%     \multirow{2}{*}{Location} &
%       \multicolumn{2}{c}{Gender} &
%       \multicolumn{3}{c}{Age} \\
%       & {Female} & {Male} & {U35} & {BET} & {O45}  \\
%       \midrule
%     Denmark & 396 $($ 4.35 $)$ & 1229 $($ 4.09 $)$ & 692 $($ 4.21 $)$ & 402 $($ 4.07 $)$ & 531 $($ 4.15 $)$  \\
%     France & 226 $($ 4.38 $)$ & 626 $($ 4.37 $)$ & 378 $ ($ 4.44 $)$ & 137 $($ 4.33 $)$ & 337 $($ 4.31 $)$  \\
%     Germany & 181 $($ 4.44 $)$ & 573 $($ 4.32 $)$ & 459 $($ 4.36 $)$ & 121 $($ 4.01 $)$ & 174 $($ 4.48 $)$  \\
%     UK & 56,218 $($4.38 $)$ & 84,409 $($4.31$)$ & 35,973 $($4.31 $)$ & 27,864 $($4.24$)$ & 76,790 $($4.44$)$  \\
%     US & 15,831$($4.50$)$ & 24,694 $($4.54$)$ & 17,986 $($4.54$)$  & 7,182$($4.48$)$ &  15,357 $($4.52$)$ \\
%     \bottomrule
%   \end{tabular}
% \end{table*}

% Given a user-generated text review, the goal is to generate a latent representation which preserves the semantic meaning of the text for sentiment analysis task while protects private information of the user, age, gender, and location. We perform 10-fold cross validation to evaluate the effectiveness of the proposed model with respect to both given task and privacy perspective.
\subsubsection{Model and Parameter Settings}
For the document auto-encoder $A$, we use single-layer RNN with GRU cell of input/hidden dimension with $d$=64. For semantic and private attribute discriminators, we use feed-forward networks with single hidden layer with the dimension of hidden state set as 200, and a sigmoid output layer, which is determined through grid search. The parameters $\alpha$ and $\lambda$ are determined through cross-validation, and are set as $\alpha = 1$ and $\lambda=0.01$. The upper-bound constraint $c_1$ for the value of parameter $\epsilon$ is also set as $c_1= 0.1$  to ensure the  $\epsilon$-differential privacy, $\epsilon=0.1$ for the learned representation. %HL Recall that the smaller the value of $\epsilon$ is, the smaller the privacy loss will be. %We thus select a small value of upper-bound constraint $c_1 = 0.1$. 
%Note that this ensures at least the $\epsilon$-differential privacy, $\epsilon=0.1$ for the learned representation.
Note that exploring the best sentiment predictor is not the focus of our work and it can be easily replaced by different models designed for this task.
\subsection{Task 2: Part-of-speech (POS) Tagging}
POS tagging is another language processing application which is framed as a sequence tagging problem~\cite{hovy2015user}. %Next, we describe the used dataset and model for this task.
\subsubsection{Data}
For this task we use a manually POS tagged version of TrustPilot dataset in English. This data is obtained from Hovy et al.~\cite{hovy2015tagging} and consists of 600 sentences, each tagged with POS information based on the Google Universal POS tagset~\cite{petrov2012universal} and also labeled with both gender and age of the users. The gender attribute is categorized into male and female, and age attribute is categorized into two groups over-45, under-35. We follow the setting of~\cite{li2018towards} and use Web English Tree-bank (WebEng)~\cite{bies2012english} as a pre-training tagging model because of the small quantity of text available for this task. WebEng is similar to TrustPilot datasets w.r.t. the domain as both contains unedited user generated textual data.
\subsubsection{Model and Parameter Settings}
Similar to the sentiment analysis task, we use single-layer RNN with GRU cell of input/hidden dimension with $d$=64 for document auto-encoder $A$. For semantic discriminator (i.e., POS tag predictor), we use bi-directional LSTM:
\begin{align}
&\mathbf{h}_i = LSTM(x^i, \mathbf{h}_{i-1}; \theta_h), \quad \mathbf{h}'_i = LSTM(x^i, \mathbf{h}'_{i+1}; \theta'_h) \\ \nonumber
%& \mathbf{h}'_i = LSTM(x_i, \mathbf{h}'_{i+1}; \theta'_h) \\ \nonumber
& y_i = Categorical (\phi ([\mathbf{h}_i;\mathbf{h}'_i]);\theta_0)
\end{align}
where $x^i|_{i=1}^m$ is the input sequence with $m$ words, $\mathbf{h}_i$ is the $i$-th hidden state, $\mathbf{h}_0$ and $\mathbf{h}'_{m+1}$ are terminal hidden states set to zero, $[.;.]$ denotes vectors concatenation and $\phi$ is a linear transformation. The dimension of the hidden layer is set as 200. We apply a dropout rate of 0.5 to all hidden layers during training.%$[.;.]$ denotes vectors concatenation, $x_i$, $i = 1, 2, ..N$ the input sequence, $\mathbf{h}_i$, the $i$-th hidden state and $h_0$ and $h'_{N+1}$ are the terminal hidden states set to zero, and $\phi$ a linear transformation. The dimension of the hidden layer is set as 200.

For the private attribute discriminator, we use feed-forward networks with single hidden layer with the dimension of hidden state set as 200, and a sigmoid output layer (determined via grid search). The input to this network is final hidden representation $[\mathbf{h}_m;\mathbf{h}'_0]$. For hyperparameters, we set values of $\alpha$ and $\lambda$ as $\alpha = 1$ and $\lambda=0.01$ which are determined through cross-validation. The upper-bound constraint for the value of $\epsilon$ is also set as $c_1= 0.1$. %HL Note that the smaller the value of $\epsilon$, the smaller the privacy loss will be. %We thus select a small value of upper-bound constraint $c_1 = 0.1$. 
% \textbf{Parameter Settings.} For the document auto-encoder $A$, we use single-layer RNN with GRU cell of input/hidden dimension with $d$=64. For semantic and private attribute discriminators, we use feed-forward networks with single hidden layer with the dimension of hidden state set as 200, and a sigmoid output layer. The parameters $\alpha$ and $\lambda$ are determined through cross-validation, and are set as $\alpha = 1$ and $\lambda=0.01$. The upper-bound constraint $c_1$ for the value of parameter $\epsilon$ is also set as $c_1= 0.1$. Recall that the smaller the value of $\epsilon$, the smaller the privacy loss will be. %We thus select a small value of upper-bound constraint $c_1 = 0.1$.
% Note that this ensures at least the $\epsilon$-differential privacy, $\epsilon=0.1$ for the learned representation.
Note that exploring the best POS tagger is not the focus of our work and it can be easily replaced by different models designed for this task.
%the models used for POS tagger and private attribute discriminator can be easily replaced by different models designed for these tasks.
\subsection{Experimental Design}
We perform 10-fold cross validation for POS tagging and sentiment analysis tasks. We follow state-of-the-art research and report accuracy score to evaluate the utility of the generated data for the given POS tagging~\cite{brants2000tnt,hovy2015tagging} or sentiment analysis task~\cite{dos2014deep}.
%To evaluate the utility of the generated data for the given task, we report test performance in terms of the accuracy score.
 In particular, for the sentiment prediction task, we report accuracy for correctly predicting rating of reviews. We also report tagging accuracy for POS tagging task. To examine the text representation in terms of obscuring private attributes, we report test performance in terms of $F1$ score for predicting private attributes. Note that the private attributes for sentiment task include age, gender and location while private attributes for tagging task include gender and age.

We compare \textsc{DPText} in both tasks with the following baselines:
\begin{itemize}[leftmargin=*]
	\item \textbf{\textsc{Original}}: This is a variant of \textsc{DPText} and publishes the original representation $\textbf{z}$ without adding noise or utilizing $D_S$ and $D_P$ discriminators.
    \item \textbf{\textsc{DifPriv}}: This baseline adds Laplacian noise to the original representation $\textbf{z}$ according to Eq. \ref{Laplacian} (i.e., $Lap(\frac{\Delta}{\epsilon})$, $\epsilon = 0.1$, $\Delta = 2d$) without utilizing $D_S$ and $D_P$ discriminators. Note that this method makes the final representation $\epsilon$-differentially private. We compare our model against this method to investigate the effectiveness of semantic and private attribute discriminators.
	%\item \textbf{Simple-2}: This is a variation of \textsc{OurModel} and publishes the representation when only semantic meaning discriminator $\mathcal{D}_S$ is utilized.
	\item \textbf{\textsc{ADV-ALL}}~\cite{li2018towards}: This method utilizes the idea of adversarial learning and has two components, generator, discriminator. It generates a text representation that has high quality for the given task but has poor quality for inference of private attributes. %We compare our model against this method to see how well adding optimal value of noise can preserve privacy in practice
\end{itemize}
%Although the work of~\cite{li2018towards} proposes a text privacy preserving framework, we cannot compare \textsc{OurModel} with it. The reason is that \cite{li2018towards} \textit{generates} a task representation, while we \textit{manipulate} the original representation (which is derived in an unsupervised manner) by adding noise to it. The benefit of manipulating the original representation is two fold. First, the performance of our model does not depend on the process which generates the original representation, and second this representation could be generated via any model such as \textbf{.....}.

In both tasks, semantic discriminator $D_S$ is trained on the train data and applied to test data for predicting sentiment and POS tags. Similarly, we can apply private attribute discriminator $D_P$ %HL (trained on train data) to test data 
where it plays the role of an adversary trying to infer the private attributes of the user based on the textual representation. Private attribute discriminator $D_P$ is also trained on the train data and applied to test data for evaluation. Higher accuracy score for semantic discriminator $D_S$ indicates that representation has high utility for the given task, while lower $F1$ score for private attribute discriminator $D_P$ demonstrates that the textual representation has higher privacy for individuals due to obscuring their private information.
\subsection{Experimental Results}
\subsubsection{Performance Comparison}
For evaluating the quality of the learned text representation, we answer questions \textbf{Q1}, \textbf{Q2} and \textbf{Q3} for two different natural language processing tasks, i.e., sentiment prediction and POS tagging. The experimental results for different methods are demonstrated in Table~\ref{result}.%, and we have the following observations.
 
\noindent\textbf{Utility (Q1).} The results of sentiment prediction for \textsc{DPText} is comparable to the \textsc{Original} approach. This means that the representation by \textsc{DPText} preserves the semantic meaning of the textual representation according to the given task (i.e., high utility). \textsc{DifPriv} performs significantly better than \textsc{DPText} and the reason is that \textsc{DPText} applies noise at least as strong as \textsc{DifPriv} (or even more). Therefore, adding more noise results in bigger utility loss. %A\textsc{DPText} performs significantly better than \textsc{DifPriv}. This confirms the role of semantic meaning discriminator $\mathcal{D}_S$ in preserving utility and semantic meaning as it explicitly takes it into consideration when adding noise. 
 We also observe that \textsc{DPText} has better performance in terms of predicting sentiment in comparison to \textsc{ADV-ALL}. %The benefit of \textsc{DPText} over \textsc{ADV-ALL} manipulating the original representation is two fold. First, our model does not depend on the process which generates the original representation. In other words, this representation could be generated via any model such as doc2vec~\cite{le2014distributed}. Second, as we discussed in Section \ref{proof_sec} adding Laplacian noise to the text representation prevents adversary from learning the input text and minimizes re-identification of users by guaranteeing $\epsilon$-differential privacy. %This is because \textsc{ADV-ALL}~\cite{li2018towards} \textit{generates} a task-based representation, while we \textit{manipulate} the original representation (which is constructed independent of the given task) by adding noise to it. The benefit of manipulating the original representation is two fold. First, our model does not depend on the process which generates the original representation. In other words, this representation could be generated via any model such as doc2vec~\cite{le2014distributed}. Second, as we discussed in Section \ref{proof_sec} adding Laplacian noise to the text representation prevents adversary from learning the input text and minimizes re-identification of users by guaranteeing $\epsilon$-differential privacy. 

The accuracy of POS tagging task is higher when \textsc{DPText} is utilized rather than when \textsc{Original} is used. This is because POS tagging results are biased toward gender, age and location~\cite{hovy2015tagging,jorgensen2016learning}. In other words, this information affects the performance of tagging task. Removing private information from the latent representation results in removing this type of bias for tagging task. Therefore, the learned representation is more robust and results in a more accurate tagging. \textsc{DPText} also has better performance than \textsc{DifPriv} due to removal of private information and thus bias. Besides, results demonstrate that \textsc{DPText} outperforms \textsc{ADV-ALL}. These results indicate the effectiveness of \textsc{DPText} in preserving semantic meaning of the learned text representation. 
%Besides, \textsc{DPText} has significantly better performance w.r.t obscuring private information in comparison to \textsc{ADV-ALL}. The results for sentiment task with \textsc{DPText} is also comparable with \textsc{ADV-ALL}.

%These results confirm the effectiveness of \textsc{DPText} in preserving the text semantic meaning given the specific task.
\begin{table}[t]\vspace{-13pt}
%\small
	\subfloat[\bf{Sentiment Prediction Task}]{
	\begin{tabular}{lp{1cm}p{1cm}p{1cm}p{1cm}}%lS
		\toprule
		\multirow{2}{*}{Model} &
		\multicolumn{1}{c}{Sentiment} &
		\multicolumn{3}{c}{Private Attribute (F1) } \\
		& (Acc) & {Age} & {Loc} & {Gen}   \\
		\midrule
		\textbf{\textsc{Original}} & 0.7493  & 0.3449  & 0.1539  & 0.5301   \\ 
        \textbf{\textsc{DifPriv}} & 0.7397 & 0.3177& 0.1411& 0.5118\\
		
		\textbf{\textsc{ADV-ALL}} & 0.7165 & 0.3076  & 0.1080  & 0.4716 \\
		%    $\textsc{DPText}_{16}$ & \textbf{0.7586} & 0.2688  & 0.1476   & 0.4317 \\
	%	\textbf{\textsc{DPText}} & \textbf{0.7318} & \textbf{0.1994}  & \textbf{0.0581}   & \textbf{0.3911} \\
		\textbf{\textsc{DPText}} & 0.7318 & 0.1994  & 0.0581   & 0.3911 \\
		\bottomrule
	\end{tabular}}
% 	\caption{\textbf{Accuracy score over the sentiment prediction and $F1$ score for private attribute prediction tasks.}}\label{result}%\vspace{-0.5cm}
%\end{table}

%\begin{table}[t]
\quad
		\subfloat[\bf{POS Tagging Task}]{
	\begin{tabular}{lp{1cm}p{1cm}p{1cm}p{1cm}}%lS
		\toprule
		\multirow{2}{*}{Model} &
		\multicolumn{1}{c}{POS Tagging} &
		\multicolumn{2}{c}{Private Attribute (F1) } \\
		& (Acc) & {Age} & {Gen}   \\
		\midrule
		\textbf{\textsc{Original}} & 0.8913  & 0.4018  &  0.5627   \\ 
        \textbf{\textsc{DifPriv}} &0.8982 & 0.3911 & 0.5417\\
		
		\textbf{\textsc{ADV-ALL}} & 0.8901 & 0.3514  &0.5008 \\
		%    $\textsc{DPText}_{16}$ & \textbf{0.7586} & 0.2688  & 0.1476   & 0.4317 \\
		%\textbf{\textsc{DPText}} & \textbf{0.9257} & \textbf{0.2218}  &  \textbf{0.3865} \\
		\textbf{\textsc{DPText}} & 0.9257 & 0.2218  &  0.3865 \\
		\bottomrule
	\end{tabular}}\vspace{-8pt}
	\caption{\textbf{Accuracy for sentiment prediction and POS tagging and $F1$ for evaluating private attribute prediction task.}}\label{result}\vspace{-0.8cm}
% 	\caption{\textbf{Accuracy score for two different natural language processing tasks, i.e., sentiment prediction and POS tagging. $F1$ score is also used to evaluate private attribute prediction task. Higher accuracy values show higher utility, while lower $F1$ score values indicate higher privacy.}}\label{result}\vspace{-0.75cm}
\end{table}
\begin{table}[t]\vspace{-15pt}
%\small
	\subfloat[\textbf{Sentiment Prediction Task}]{
	\begin{tabular}{lp{1cm}p{1cm}p{1cm}p{1cm}}%lS
		
		\toprule
		\multirow{2}{*}{Model} &
		\multicolumn{1}{c}{Sentiment} &
		\multicolumn{3}{c}{Private Attribute (F1) } \\
		& (Acc) & {Age} & {Loc} & {Gen}   \\
		\midrule
		\textbf{\textsc{DPText}} & 0.7318 & 0.1994  & 0.0581   & 0.3911 \\ \midrule
		\textbf{\textsc{DPTextAge}} & 0.7573 & 0.2248  & 0.1012  & 0.3982  \\   
		\textbf{\textsc{DPTextLoc}} & 0.7360 & 0.2861  & 0.0731  & 0.4100  \\
		\textbf{\textsc{DPTextGen}} & 0.7347 & 0.2997  & 0.0623  & 0.4053  \\
		
		\bottomrule
	\end{tabular}}
% 	\caption{\textbf{Impact of different private attribute discriminators on \textsc{DPText} for sentiment prediction task.}}\label{result2}
% \end{table}

% \begin{table}[t]
	\subfloat[\textbf{POS Tagging Task}]{
	\begin{tabular}{lp{1cm}p{1cm}p{1cm}p{1cm}}%lS
		
		\toprule
		\multirow{2}{*}{Model} &
		\multicolumn{1}{c}{POS Tagging} &
		\multicolumn{2}{c}{Private Attribute (F1) } \\
		& (Acc) & {Age} &  {Gen}   \\
		\midrule
		\textbf{\textsc{DPText}} & 0.9257 & 0.2218  & 0.3865 \\ \midrule
		\textbf{\textsc{DPTextAge}} & 0.9218 & 0.2111   & 0.4179  \\   
		\textbf{\textsc{DPTextGen}} & 0.9361 & 0.2412   & 0.3916  \\
		
		\bottomrule
	\end{tabular}}\vspace{-5pt}
	\caption{\textbf{Impact of different private attribute discriminators on \textsc{DPText} for sentiment prediction and POS tagging tasks.}}\label{result2}\vspace{-0.8cm}% Higher accuracy values show higher utility, while lower $F1$ score values indicate higher privacy.}}\label{result2}\vspace{-0.75cm}
\end{table}
\noindent \textbf{Privacy (Q2).} In the sentiment prediction task, \textsc{DPText} has significantly lower $F1$ score in comparison to \textsc{Original} and thus outperforms \textsc{Original} in terms of obscuring private information. \textsc{DPText} has significantly better performance in hiding private information than \textsc{DifPriv}. This indicates that solely adding noise and satisfying $\epsilon$-differential privacy does not protect textual information against leakage of private attributes. This further demonstrates the importance of private attribute discriminator $D_P$ in obscuring users' private information. We also observe that the learned textual representation via \textsc{DPText} hides more private information than \textsc{ADV-ALL} (lower $F1$ score). These results indicate that \textsc{DPText} can successfully obscure private information. % (i.e., high privacy). 
%Another interesting finding is when we set $\alpha=16$ in \textsc{DPText}, i.e. $\textsc{DPText}_{16}$. The results show that $\textsc{DPText}_{16}$ can beat ADV-ALL in terms of both privacy and utility which confirms the ability of our model in retaining high utility and still obscuring private information.

In the POS tagging task, $F1$ scores of \textsc{DPText} are significantly lower than \textsc{Original} approach. These results demonstrate the effectiveness of \textsc{DPText} in obscuring users' private attribute. Similarly, comparing $F1$ scores of \textsc{DPText} and \textsc{DifPriv} shows that \textsc{DPText} contains less private attribute information. This confirms the incapability of \textsc{DifPriv} in obscuring users' private information, and clearly shows the effectiveness of private attribute discriminator $D_P$. Moreover, \textsc{DPText} outperforms \textsc{ADV-ALL} method in terms of hiding user's age and gender information. It confirms that the learned textual latent representation by \textsc{DPText} preserves privacy by eliminating their sensitive information w.r.t. POS tagging task. % (i.e., high privacy).

%Our results confirm the effectiveness of \textsc{DPText} in obscuring private attribute information from the learned text representation.
\noindent \textbf{Utility-Privacy Relation (Q3).} For the sentiment prediction task, \textsc{DPText} has achieved the highest accuracy and thus reached the highest utility in comparison to other methods. It also has comparable utility results to \textsc{Original}. However, \textsc{Original} utility is preserved at the expense of significant privacy loss. Moreover, although \textsc{DifPriv} satisfies differential privacy and its performance is comparable with \textsc{DPText} for predicting sentiment, it performs poorly in obscuring private information. \textsc{DifPriv} may provide weaker privacy guaranty comparing with \textsc{DPText} since learned $\epsilon$ in \textsc{DPText} can be smaller than $\epsilon=0.1$ in \textsc{DifPriv}. In contrast, \textsc{DPText} has significantly better (best) results in terms of privacy compared to the other approaches and also achieves the least utility loss in comparison to \textsc{ADV-ALL}. For the POS tagging task, the resultant representation from \textsc{DPText} achieves the highest utility and privacy amongst all approaches. This shows the effectiveness of \textsc{DPText} in preserving semantic meaning and obscuring private information for more accurate tagging.

% \textsc{DPText} has comparable results to \textsc{Original} and \textsc{ADV-ALL} methods in terms of preserving semantic meaning of the learned text representation. Besides, \textsc{DPText} has significantly better performance w.r.t obscuring private information in comparison to other techniques. 
The results for two natural language processing tasks indicate that \textsc{DPText} learns a textual representation that %1) prevents text reconstruction, 2) is at least $\epsilon$-differentially private ($\epsilon <c_1$, $c_1 = 0.1$), 
 (1) does not contain private information, and (2) preserves the semantic meaning of the representation for the given task. % These findings confirm the effectiveness of our proposed model in terms of both privacy and utility.
\begin{figure*}[t]
	\centering
	\small
%	\begin{flushleft}
		\subfloat[Private attribute prediction w.r.t. sentiment task (F1)]{\includegraphics[scale=0.088]{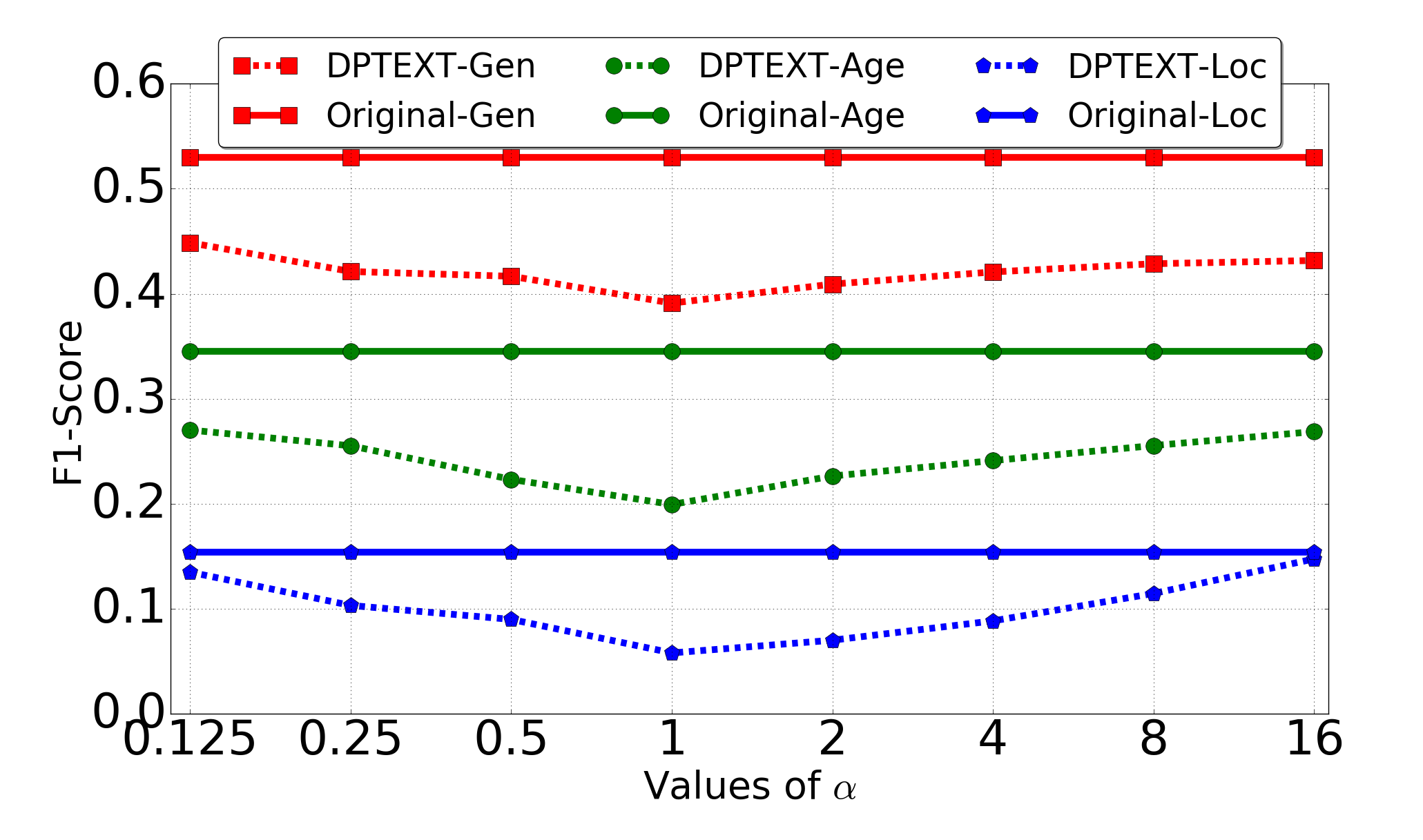}}
		\subfloat[Sentiment prediction (Acc)]{\includegraphics[scale=0.088]{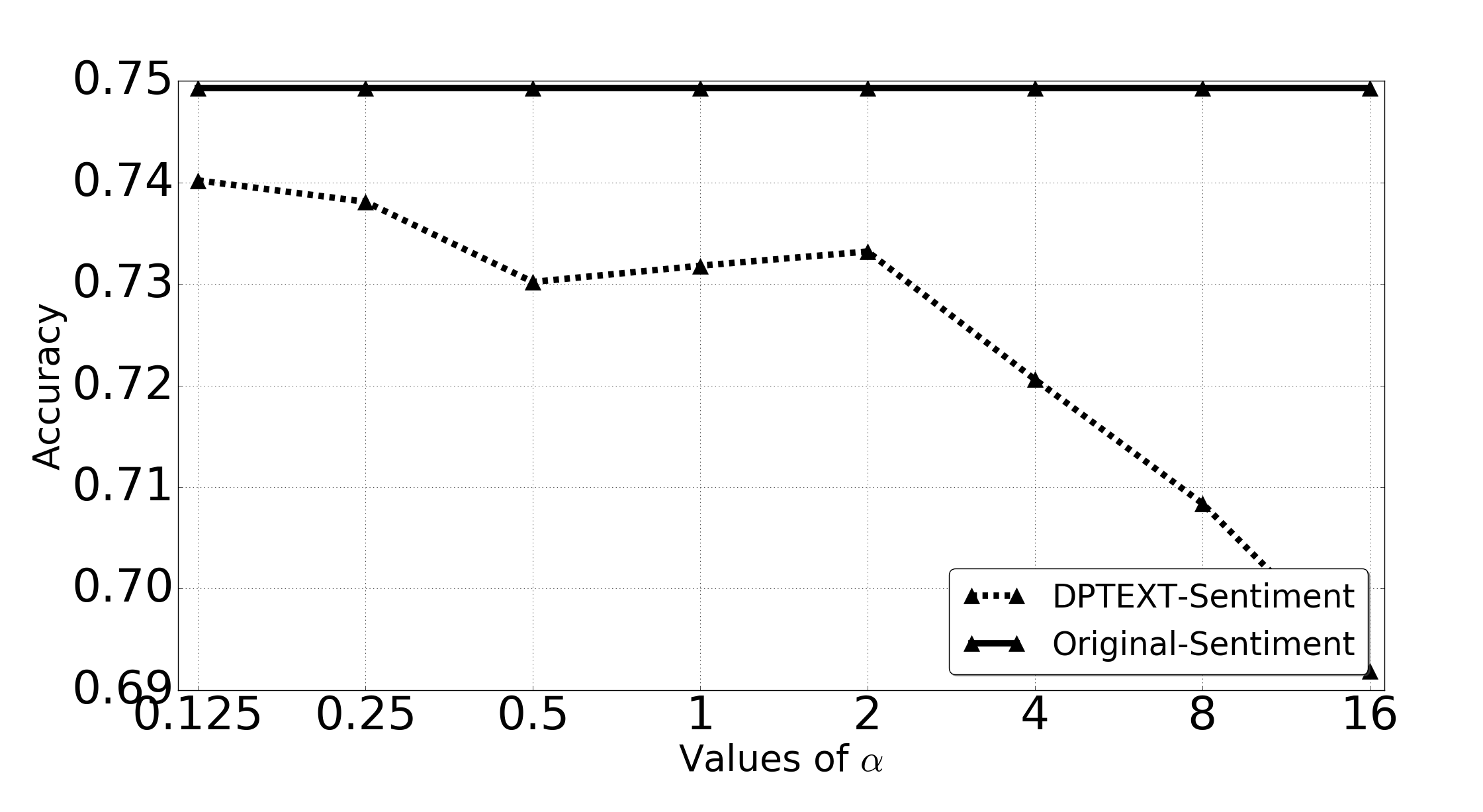}}
        \subfloat[Private attribute prediction w.r.t. pos tagging (F1)]{\includegraphics[scale=0.088]{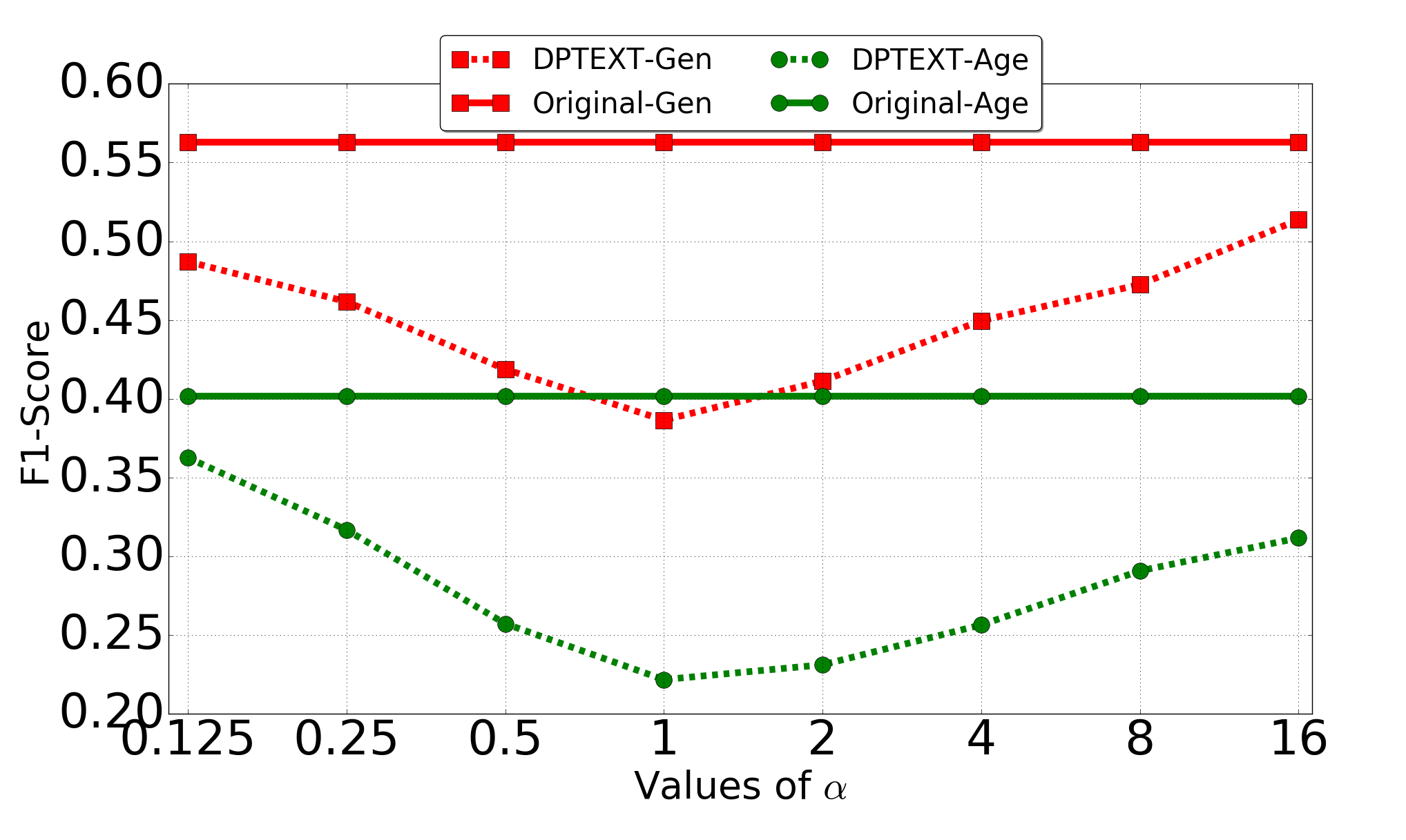}}
        \subfloat[POS tagging prediction (Acc)]{\includegraphics[scale=0.088]{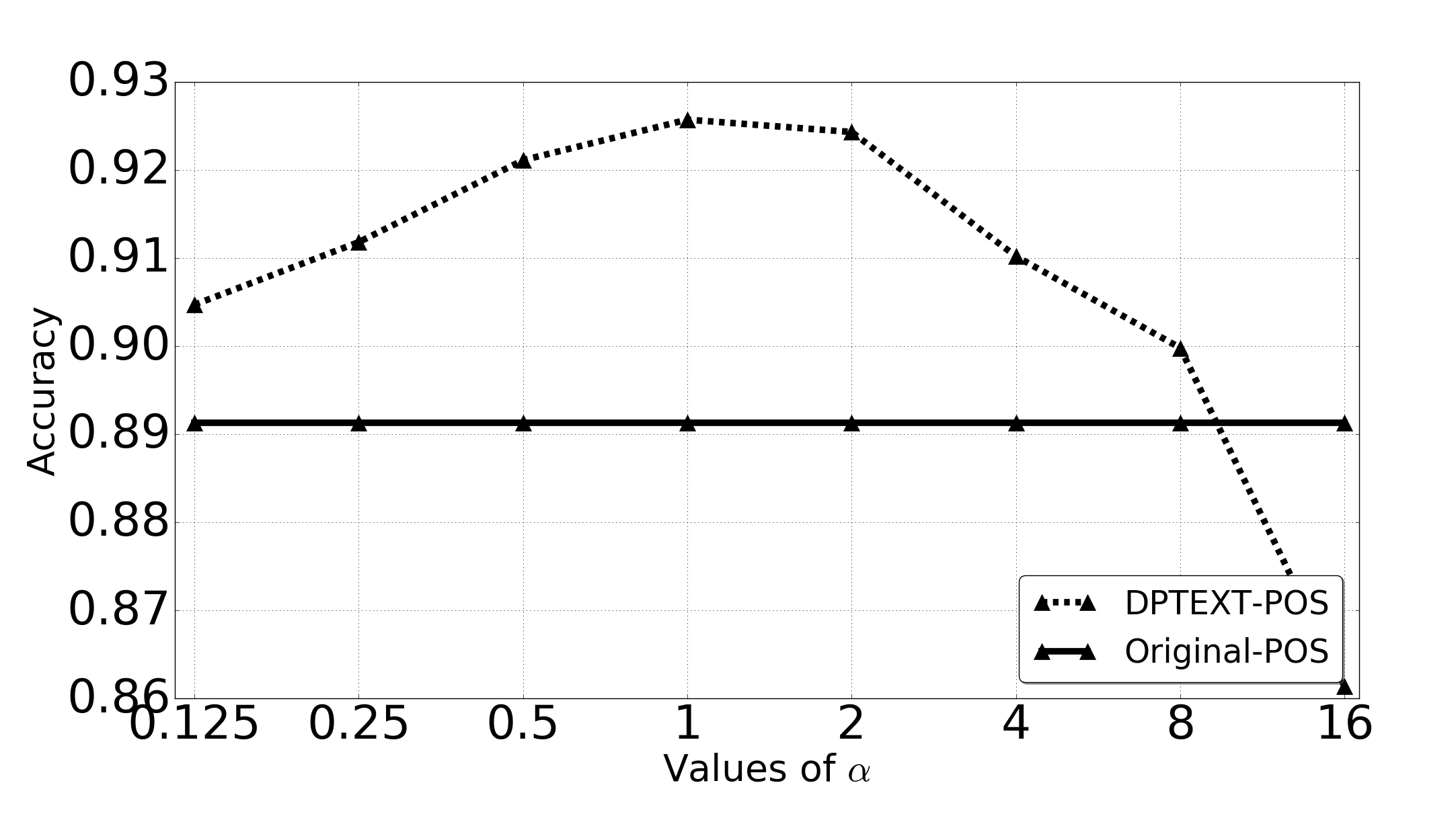}}\vspace{-5pt}
		\caption{\textbf{Performance results for private attribute and sentiment prediction tasks for different values of $\alpha$}}\label{para}
	%\end{flushleft}
	\vspace{-0.2cm}
\end{figure*}
\subsubsection{Impact of Different Components}
In this subsection, we investigate the impact of different private attribute discriminators on obscuring users' private information. To achieve this goal, we define three variants of the proposed framework, i.e., \textsc{DPText\{Age/Gen/Loc\}}. In each of these variants, the model is trained with discriminator of just one of the private attributes. For example, \textsc{DPTextAge} is trained solely with age discriminator and does not use any other private attribute discriminators during training phase. The performance comparison is shown in Table~\ref{result2}.

In sentiment prediction task, we observe that using solely one of the private attribute discriminators can result in a representation which performs better in terms of sentiment prediction, in comparison to \textsc{DPText} in which we use all three private attributes discriminators (i.e., higher utility). However, these variants perform poorly in terms of obscuring private attributes in comparison to the original \textsc{DPText} model. These results indicate that although using one discriminator in the training process can help in preserving more semantic, it can compromise the effectiveness of learned representation in obscuring attributes.

In the POS tagging task, results show that \textsc{DPText} achieves the best performance in tagging task (i.e., higher utility) in comparison to other methods that solely use one of the private attribute discriminators. The reason is that presence of age and gender related information in the text can negatively affect the tagging performance due to existing bias~\cite{hovy2015tagging,jorgensen2016learning}. \textsc{DPText} is thus more effective in removing this bias and leads to more accurate tagging in comparison to \textsc{DPTextAge} and \textsc{DPTextGen}. Similar to sentiment prediction task, we observe that \textsc{DPTextGen} with only gender attribute discriminator is less effective than \textsc{DPText} in terms of hiding private attributes information. \textsc{DPTextAge} however, has the best results in terms of obscuring age attribute information.
\vspace{-5pt}
\subsubsection{Parameter Analysis}
% \begin{figure*}[ht]
% 	\centering
% 	\small
% %	\begin{flushleft}
% 		\subfloat[Private attribute prediction (F1)]{\includegraphics[scale=0.13]{figure_1_privattr.png}}
% 		\subfloat[Sentiment prediction (Acc)]{\includegraphics[scale=0.13]{figure_1_sent.png}}
% 		\caption{\textbf{Performance results for private attribute and sentiment prediction tasks for different values of $\alpha$}}\label{para}
% 	%\end{flushleft}
% 	% \vspace{-1cm}
% \end{figure*}

% \begin{figure*}[ht]
% 	\centering
% 	\small
% 	%\begin{flushleft}
% 		\subfloat[Private attribute prediction (F1)]{\includegraphics[scale=0.13]{figure_2_privattr.png}} %0.17 before
% 		\subfloat[POS tagging prediction (Acc)]{\includegraphics[scale=0.13]{figure_2_pos.png}}
% 		\caption{\textbf{Performance results for private attribute and POS tagging tasks for different values of $\alpha$}}\label{para-pos}
% 	%\end{flushleft}
% 	% \vspace{-1cm}
% \end{figure*}
\textsc{DPText} has one important parameter $\alpha$ which controls the contribution from private attribute discriminator $D_P$. We investigate the effect of this parameter by varying it as $\{0.125, 0.25, 0.5, 1, 2, 4, 8, 16\}$. \textsc{Original-\{Age/Gen/Loc\}} shows the results for the corresponding task when the original text representation has been utilized. Results are shown in the Fig.~\ref{para}.(a-b) and Fig.~\ref{para}.(c-d) for sentiment prediction and POS tagging, respectively.

Although $\alpha$ controls the contribution of private attribute discriminator, we surprisingly observe that in both sentiment prediction and POS tagging tasks with the increase of $\alpha$, the $F1$ scores for prediction of different private attributes decrease at first up to the point that $\alpha = 1$ and then it increases. This means that the private attributes were obscured more accurately at the beginning with the increase of $\alpha$ and less later. Moreover, with the increase of $\alpha$, the accuracy of sentiment prediction task decreases. This shows that increasing the contribution of private attribute discriminator lead to decrease in the utility of resultant text representation. In case of POS tagging, the accuracy first increases and then decreases after $\alpha = 1$. This shows that removing the age and gender attributes related information results in removing the bias from learned text representation and improve the tagging task. However, after $\alpha = 1$ the utility of resultant representation decreases. Those patterns are useful for selecting the value of parameter $\alpha$ in practice.

Moreover, in both tasks, setting $\alpha=0.125$ results in an improvement in terms of the amount of hidden private information in comparison to the results of using \textsc{Original} representation. This observation supports the importance of the private attribute discriminator. Another observation is that, after $\alpha = 1$, continuously increasing $\alpha$ degrades the performance of hiding private attributes (i.e., increasing $F1$ scores) in both sentiment prediction and POS tagging tasks. %while improves the sentiment prediction task (i.e. preserving semantic meaning of the learned representation). 
 This is because the model could overfit by increasing $\alpha$ which lead to an inaccurate learned text representation in terms of preserving private attributes and semantic meaning of the text.\vspace{-5pt}
\section{Related Work}

Explosive growth of the Web not only has drastically changed the way people conduct activities and acquire information, but also has raised numerous challenges~\cite{beigi2019signed,beigi2018similar} including security~\cite{alvari2018early,alvari2017semi,alvari2019less,alvari2019hawkes} and privacy~\cite{beigi2019protecting,beigi2018securing,BeigiSigweb} issues for them. Identifying and mitigating user privacy issues has been studied from different aspects on the Web and social media (for a comprehensive survey see~\cite{beigi2018privacy}). Our work is related to a number of research which we discuss below while highlighting the differences between our work and them.

\noindent\textbf{Differential Privacy Application in Social Media.}
Differential privacy has been used for many privacy preserving applications. For example, $\epsilon$-differential privacy has been used to preserve privacy in graph data~\cite{xiao2014differentially}. %Sala et al.~\cite{sala2011sharing} and Xiao et al.~\cite{xiao2014differentially} use $\epsilon$-differential privacy mechanism to preserve privacy in graph data. %partitions the statistical representation of the graph captured by $dK$-series (degree distributions of connected components of size $K$) into clusters and then uses $\epsilon$-differential privacy mechanism to add noise to the representation in each cluster. 
%In another work Xiao et al.~\cite{xiao2014differentially} propose an anonymization approach for graph data which satisfies edge $\epsilon$-differential privacy to hide each user's connections to other users. %In particular, their method learns how to transform edges to connection probabilities via statistical Hierarchal Random Graphs under differential privacy.
%propose anonymization approaches that satisfy $\epsilon$-differential privacy in graph data by hiding each user's connections to other users.% In particular, their method partitions the statistical representation of the graph captured by $dK$-series (degree distributions of connected components of size $K$) into clusters and then uses $\epsilon$-differential privacy mechanism to add noise to the representation in each cluster. In another work, Xiao et al.~\cite{xiao2014differentially} propose an anonymization approach for graph data which satisfies edge $\epsilon$-differential privacy to hide each user's connections to other users. In particular, their method learns how to transform edges to connection probabilities via statistical Hierarchal Random Graphs under differential privacy.
Another application of differential privacy is in recommendation systems that it is utilized to construct private covariance matrices~\cite{mcsherry2009differentially} and private users' sensitive ratings~\cite{meng2018personalized}. %McSherry et al.~\cite{mcsherry2009differentially} utilize differential privacy to construct private covariance matrices so that drawing inferences about original ratings is difficult. Another example in recommendation systems is the work of Machanavajjhala et al.~\cite{machanavajjhala2011personalized} which utilizes differential privacy under Laplace and Exponential mechanisms. A recent research from Meng et al.~\cite{meng2018personalized} deploys differential privacy to perturb users' sensitive ratings. %proposes a personalized privacy preserving recommender system which is given users' sensitive and non-sensitive ratings and then utilizes differential privacy to further perturb users' ratings. 

\noindent\textbf{Privacy Preserving Web Search.}
%A user composes a query formed by one or more keywords and sends it to the search engine. 
The search engine returns a list of web pages according to a user's query formed by one or more keywords. %These search queries are a rich source of information for user profiling. 
%Jones et. al. ~\cite{jones2007know} studies the potential vulnerabilities of the search engine query logs for the first time by proposing an attack which infers a user's private information (e.g., age, gender and zip code) from query logs using classification approaches.
Privacy preserving web search approaches focus on anonymizing users search queries. %one group of works~\cite{kumar2007anonymizing,adar2007user} proposed an ad-hoc token based approach for query log anonymization in which each query string is tokenized and then each token is hashed into an identifier.
 One group of works focused on the protection of post-hoc logs~\cite{korolova2009releasing,gotz2012publishing,zhang2016anonymizing}. Korolova et al.~\cite{korolova2009releasing} releases a $(\epsilon, \delta)$-differential private query click graph. %This approach exploits differential privacy to identify which queries could be released. Then it adds Laplacian noise to the number of users who clicked on a link. 
The work of Zhang et al.~\cite{zhang2016anonymizing} makes a significant improvement over~\cite{korolova2009releasing} by providing an $(\epsilon)$-differential privacy. Another set of works includes client-side ones focuses on search query obfuscation~\cite{gervais2014quantifying,howe2009trackmenot} which adds dummy search queries (collected from popular websites and searched query terms) on behalf of users.% which are collected from popular websites and popular recently searched query terms. These approaches are user-centric and do not need cooperation of the engine. 
Beigi et al.~\cite{beigi2019protecting} also proposes a method for preserving web browsing history by inferring how many and what links should be added to a user's browsing history to preserve his privacy while retaining the utility.

Preserving privacy of text is more challenging than web search privacy preserving since queries often include few keywords while text data consists of much larger numbers of words.

\noindent\textbf{Textual Data Anonymization.}
%There are various sources of user-generated textual information such as posts on social media, reviews, and emails and publishing this information would pose similar privacy issues. 
Few works consider the privacy of textual user data~\cite{hakkini2006sanitization,anandan2012t,TextAnonymization,li2018towards}. The work of~\cite{hakkini2006sanitization} introduces possible privacy threats of document repositories, 1) name entity recognition, and 2) author identification. It then introduces the concept of $k$-author anonymity to address the latter issue. However, this work failed to provide technical solutions to address the privacy challenges. Another work from Anandan et al.~\cite{anandan2012t} studies removing PII from text. It first introduces $t$-Plausibility notion and then propose information theoretic based algorithms which select and generalize sensitive keywords to satisfy $t$-Plausibility.
%. Then, it proposes different information theoretic based algorithms which select and generalize sensitive keywords in the given text to satisfy $t$-Plausibility. 
Its is that it does not address textual representation re-identification and removal of hidden private information.  %had ideas but didn't back them up with any technical solutions.
%The work of~\cite{TextAnonymization} first introduces a verified version of differential privacy specified for textual data, namely, $\epsilon$-Text Indistinguishability to overcome the curse of dimensionality problem when original differential privacy is deployed on high-dimensional textual data. It then proposes a framework which perturbs user-keyword matrix by adding Laplacian noise to satisfy $\epsilon$-Text Indistinguishability. 
 The work of Zhang et al.~\cite{TextAnonymization} introduces a verified version of differential privacy specified for textual data to overcome the curse of dimensionality problem, namely, $\epsilon$-TextIndistinguishability and satisfy it by adding Laplacian noise. Another work~\cite{li2018towards} uses the idea of adversarial learning to generate text representation. Their framework consists of a generator which generates representation w.r.t. given task and a discriminator which ensures the representation does not contain private information. 

Our work is different from~\cite{TextAnonymization,li2018towards}. First, \cite{TextAnonymization} does not consider the task that given text will be used for. This results in a representation which lacks utility in practice. Moreover, this framework does not handle leakage of private attribute. However, the semantic and private attribute discriminators in \textsc{DPText} ensure the utility and privacy. Second, in \cite{li2018towards} the input textual information could be recovered from the representation if the adversary has preliminary knowledge of the training model and this could be done easily through reverse engineering by a GAN attack algorithm~\cite{hitaj2017deep}. Moreover, \cite{li2018towards} does not consider the risk of text representation re-identification, while \textsc{DPText} does. \textsc{DPText} does not depend on the process of generating original representation and this representation could be generated via any model such as doc2vec~\cite{le2014distributed}. %\textsc{DPText} learns a representation that preserves semantic meaning, is $\epsilon$-differentially private and is prune to the mentioned issues.
\vspace{-0.2cm}
\section{Conclusion}
In this paper, we propose a double privacy preserving text representation learning framework, \textsc{DPText}, which learns a text representation that (1) is differentially private, (2) obscures users' private information, and (3) retains high utility for a given task. It has four main components, 1) an auto-encoder, 2) differential-privacy-based noise adder, 3) a semantic meaning discriminator, and 4) a private attribute discriminator. Our theoretical and empirical results shows the effectiveness of \textsc{DPText} in minimizing chances of learned textual representation re-identification, obscuring private attribute information and preserving semantic meaning of the text. 
% The proposed framework is trained via adversarial learning.%\textsc{DPText} is trained via adversarial learning. 
% Our theoretical findings confirm the fact that \textsc{DPText} minimizes the chance of re-identifying users in textual data. Evaluation on real-world data for two natural language processing tasks, i.e. sentiment prediction and part-of-speech tagging, shows the effectiveness of \textsc{DPText} in obscuring private attribute information while preserving semantic meaning of the text. 
 One future direction for this work is to generate privacy preserving text (e.g., sentences, paragraphs) rather than laten representation which is critical for having interpretable results. 
We also adopt Laplacian noise rather than Guassian noise as it provides stronger guarantees for differential privacy. Another future direction is to adopt Gaussian noise to satisfy differential privacy and further examine how it affects obscuring private attribute information and preserving semantic meaning of data. It would also be interesting to generalize the proposed model for other types of data, e.g., location traces.
 
 %There are several interesting directions for further investigations. 
% In this work, we learned the latent text representation. One future direction is to generate privacy preserving text (e.g., sentences, paragraphs) which is critical for having interpretable results. 
% We also adopt Laplacian noise rather than Guassian noise as it provides stronger guarantees for differential privacy. Another future direction is to adopt Gaussian noise to satisfy differential privacy and further examine how adopting Guassian noise may affect preserving users' private attribute information and semantic meaning of the textual data. It would also be interesting to generalize the proposed model for preserving privacy of other types of user data, e.g., user-item interactions in product review websites, and location traces.

\begin{acks}
This material is based upon the work supported, in part, by NSF \#1614576, ARO W911NF-15-1-0328 and ONR N00014-17-1-2605.
\end{acks}

\bibliographystyle{ACM-Reference-Format}
%\bibliography{sample-bibliography}

\begin{thebibliography}{51}
	

	
	\bibitem[\protect\citeauthoryear{??}{bbc}{[n. d.]}]%
	{bbc}
	\bibinfo{year}{[n. d.]}\natexlab{}.
	\newblock \bibinfo{title}{How does Twitter make money?}
	\newblock
	\bibinfo{howpublished}{\url{https://www.bbc.com/news/business-24397472}}.
	\newblock
	\newblock
	\shownote{Accessed: 2013-11-07.}
	
	
	\bibitem[\protect\citeauthoryear{Alvari, Shaabani, Sarkar, Beigi, and
		Shakarian}{Alvari et~al\mbox{.}}{2019}]%
	{alvari2019less}
	\bibfield{author}{\bibinfo{person}{Hamidreza Alvari}, \bibinfo{person}{Elham
			Shaabani}, \bibinfo{person}{Soumajyoti Sarkar}, \bibinfo{person}{Ghazaleh
			Beigi}, {and} \bibinfo{person}{Paulo Shakarian}.}
	\bibinfo{year}{2019}\natexlab{}.
	\newblock \showarticletitle{Less is More: Semi-Supervised Causal Inference for
		Detecting Pathogenic Users in Social Media}. In
	\bibinfo{booktitle}{\emph{Companion Proceedings of The 2019 World Wide Web
			Conference}}. ACM, \bibinfo{pages}{154--161}.
	\newblock
	
	
	\bibitem[\protect\citeauthoryear{Alvari, Shaabani, and Shakarian}{Alvari
		et~al\mbox{.}}{2018}]%
	{alvari2018early}
	\bibfield{author}{\bibinfo{person}{Hamidreza Alvari}, \bibinfo{person}{Elham
			Shaabani}, {and} \bibinfo{person}{Paulo Shakarian}.}
	\bibinfo{year}{2018}\natexlab{}.
	\newblock \showarticletitle{Early Identification of Pathogenic Social Media
		Accounts}. In \bibinfo{booktitle}{\emph{IEEE Intelligence and Security
			Informatics (ISI)}}. IEEE.
	\newblock
	
	
	\bibitem[\protect\citeauthoryear{Alvari and Shakarian}{Alvari and
		Shakarian}{2019}]%
	{alvari2019hawkes}
	\bibfield{author}{\bibinfo{person}{Hamidreza Alvari} {and}
		\bibinfo{person}{Paulo Shakarian}.} \bibinfo{year}{2019}\natexlab{}.
	\newblock \showarticletitle{Hawkes Process for Understanding the Influence of
		Pathogenic Social Media Accounts}. In \bibinfo{booktitle}{\emph{2019 2nd
			International Conference on Data Intelligence and Security (ICDIS)}}. IEEE.
	\newblock
	
	
	\bibitem[\protect\citeauthoryear{Alvari, Shakarian, and Snyder}{Alvari
		et~al\mbox{.}}{2017}]%
	{alvari2017semi}
	\bibfield{author}{\bibinfo{person}{Hamidreza Alvari}, \bibinfo{person}{Paulo
			Shakarian}, {and} \bibinfo{person}{JE~Kelly Snyder}.}
	\bibinfo{year}{2017}\natexlab{}.
	\newblock \showarticletitle{Semi-supervised learning for detecting human
		trafficking}.
	\newblock \bibinfo{journal}{\emph{Security Informatics}} \bibinfo{volume}{6},
	\bibinfo{number}{1} (\bibinfo{year}{2017}), \bibinfo{pages}{1}.
	\newblock
	
	
	\bibitem[\protect\citeauthoryear{Anandan, Clifton, Jiang, Murugesan,
		Pastrana-Camacho, and Si}{Anandan et~al\mbox{.}}{2012}]%
	{anandan2012t}
	\bibfield{author}{\bibinfo{person}{Balamurugan Anandan}, \bibinfo{person}{Chris
			Clifton}, \bibinfo{person}{Wei Jiang}, \bibinfo{person}{Mummoorthy
			Murugesan}, \bibinfo{person}{Pedro Pastrana-Camacho}, {and}
		\bibinfo{person}{Luo Si}.} \bibinfo{year}{2012}\natexlab{}.
	\newblock \showarticletitle{t-Plausibility: Generalizing Words to Desensitize
		Text}.
	\newblock \bibinfo{journal}{\emph{Transactions on Data Privacy}}
	\bibinfo{volume}{5}, \bibinfo{number}{3} (\bibinfo{year}{2012}),
	\bibinfo{pages}{505--534}.
	\newblock
	
	
	\bibitem[\protect\citeauthoryear{Barbaro, Zeller, and Hansell}{Barbaro
		et~al\mbox{.}}{2006}]%
	{barbaro2006face}
	\bibfield{author}{\bibinfo{person}{Michael Barbaro}, \bibinfo{person}{Tom
			Zeller}, {and} \bibinfo{person}{Saul Hansell}.}
	\bibinfo{year}{2006}\natexlab{}.
	\newblock \showarticletitle{A face is exposed for AOL searcher no. 4417749}.
	\newblock \bibinfo{journal}{\emph{New York Times}} \bibinfo{volume}{9},
	\bibinfo{number}{2008} (\bibinfo{year}{2006}), \bibinfo{pages}{8}.
	\newblock
	
	
	\bibitem[\protect\citeauthoryear{Beigi, Guo, Nou, Zhang, and Liu}{Beigi
		et~al\mbox{.}}{2019a}]%
	{beigi2019protecting}
	\bibfield{author}{\bibinfo{person}{Ghazaleh Beigi}, \bibinfo{person}{Ruocheng
			Guo}, \bibinfo{person}{Alexander Nou}, \bibinfo{person}{Yanchao Zhang}, {and}
		\bibinfo{person}{Huan Liu}.} \bibinfo{year}{2019}\natexlab{a}.
	\newblock \showarticletitle{Protecting user privacy: An approach for
		untraceable web browsing history and unambiguous user profiles}. In
	\bibinfo{booktitle}{\emph{Proceedings of the Twelfth ACM International
			Conference on Web Search and Data Mining}}. ACM, \bibinfo{pages}{213--221}.
	\newblock
	
	
	\bibitem[\protect\citeauthoryear{Beigi and Liu}{Beigi and Liu}{2018a}]%
	{beigi2018privacy}
	\bibfield{author}{\bibinfo{person}{Ghazaleh Beigi} {and} \bibinfo{person}{Huan
			Liu}.} \bibinfo{year}{2018}\natexlab{a}.
	\newblock \showarticletitle{Privacy in social media: Identification, mitigation
		and applications}.
	\newblock \bibinfo{journal}{\emph{arXiv preprint arXiv:1808.02191}}
	(\bibinfo{year}{2018}).
	\newblock
	
	
	\bibitem[\protect\citeauthoryear{Beigi and Liu}{Beigi and Liu}{2018b}]%
	{beigi2018similar}
	\bibfield{author}{\bibinfo{person}{Ghazaleh Beigi} {and} \bibinfo{person}{Huan
			Liu}.} \bibinfo{year}{2018}\natexlab{b}.
	\newblock \showarticletitle{Similar but different: Exploiting users’
		congruity for recommendation systems}. In
	\bibinfo{booktitle}{\emph{International Conference on Social Computing,
			Behavioral-Cultural Modeling and Prediction and Behavior Representation in
			Modeling and Simulation}}. Springer, \bibinfo{pages}{129--140}.
	\newblock
	
	
	\bibitem[\protect\citeauthoryear{Beigi and Liu}{Beigi and Liu}{2019}]%
	{BeigiSigweb}
	\bibfield{author}{\bibinfo{person}{Ghazaleh Beigi} {and} \bibinfo{person}{Huan
			Liu}.} \bibinfo{year}{2019}\natexlab{}.
	\newblock \showarticletitle{"Identifying Novel Privacy Issues of Online Users
		on Social Media Platforms" by Ghazaleh Beigi and Huan Liu with Martin Vesely
		As Coordinator}.
	\newblock \bibinfo{journal}{\emph{SIGWEB Newsl.}} \bibinfo{number}{Winter},
	Article \bibinfo{articleno}{4} (\bibinfo{date}{Feb.} \bibinfo{year}{2019}),
	\bibinfo{numpages}{7}~pages.
	\newblock
	
	
	\bibitem[\protect\citeauthoryear{Beigi, Ranganath, and Liu}{Beigi
		et~al\mbox{.}}{2019b}]%
	{beigi2019signed}
	\bibfield{author}{\bibinfo{person}{Ghazaleh Beigi}, \bibinfo{person}{Suhas
			Ranganath}, {and} \bibinfo{person}{Huan Liu}.}
	\bibinfo{year}{2019}\natexlab{b}.
	\newblock \showarticletitle{Signed Link Prediction with Sparse Data: The Role
		of Personality Information}. In \bibinfo{booktitle}{\emph{Companion
			Proceedings of The 2019 World Wide Web Conference}}. ACM,
	\bibinfo{pages}{1270--1278}.
	\newblock
	
	
	\bibitem[\protect\citeauthoryear{Beigi, Shu, Guo, Wang, and Liu}{Beigi
		et~al\mbox{.}}{2019c}]%
	{beigi2019text}
	\bibfield{author}{\bibinfo{person}{Ghazaleh Beigi}, \bibinfo{person}{Kai Shu},
		\bibinfo{person}{Ruocheng Guo}, \bibinfo{person}{Suhang Wang}, {and}
		\bibinfo{person}{Huan Liu}.} \bibinfo{year}{2019}\natexlab{c}.
	\newblock \showarticletitle{Privacy Preserving Text Representation Learning}.
	In \bibinfo{booktitle}{\emph{Proceedings of the 30th on Hypertext and Social
			Media}} \emph{(\bibinfo{series}{HT '19})}. \bibinfo{publisher}{ACM}.
	\newblock
	
	
	\bibitem[\protect\citeauthoryear{Beigi, Shu, Zhang, and Liu}{Beigi
		et~al\mbox{.}}{2018}]%
	{beigi2018securing}
	\bibfield{author}{\bibinfo{person}{Ghazaleh Beigi}, \bibinfo{person}{Kai Shu},
		\bibinfo{person}{Yanchao Zhang}, {and} \bibinfo{person}{Huan Liu}.}
	\bibinfo{year}{2018}\natexlab{}.
	\newblock \showarticletitle{Securing social media user data: An adversarial
		approach}. In \bibinfo{booktitle}{\emph{Proceedings of the 29th on Hypertext
			and Social Media}}. ACM, \bibinfo{pages}{165--173}.
	\newblock
	
	
	\bibitem[\protect\citeauthoryear{Beretta, Maccagnola, Cribbin, and
		Messina}{Beretta et~al\mbox{.}}{2015}]%
	{beretta2015interactive}
	\bibfield{author}{\bibinfo{person}{Valentina Beretta}, \bibinfo{person}{Daniele
			Maccagnola}, \bibinfo{person}{Timothy Cribbin}, {and} \bibinfo{person}{Enza
			Messina}.} \bibinfo{year}{2015}\natexlab{}.
	\newblock \showarticletitle{An interactive method for inferring demographic
		attributes in Twitter}. In \bibinfo{booktitle}{\emph{Proceedings of the 26th
			ACM Conference on Hypertext \& Social Media}}. ACM.
	\newblock
	
	
	\bibitem[\protect\citeauthoryear{Bies, Mott, Warner, and Kulick}{Bies
		et~al\mbox{.}}{2012}]%
	{bies2012english}
	\bibfield{author}{\bibinfo{person}{Ann Bies}, \bibinfo{person}{Justin Mott},
		\bibinfo{person}{Colin Warner}, {and} \bibinfo{person}{Seth Kulick}.}
	\bibinfo{year}{2012}\natexlab{}.
	\newblock \showarticletitle{English web treebank}.
	\newblock \bibinfo{journal}{\emph{Linguistic Data Consortium, Philadelphia,
			PA}} (\bibinfo{year}{2012}).
	\newblock
	
	
	\bibitem[\protect\citeauthoryear{Bowman, Vilnis, Vinyals, Dai, Jozefowicz, and
		Bengio}{Bowman et~al\mbox{.}}{2015}]%
	{bowman2015generating}
	\bibfield{author}{\bibinfo{person}{Samuel~R Bowman}, \bibinfo{person}{Luke
			Vilnis}, \bibinfo{person}{Oriol Vinyals}, \bibinfo{person}{Andrew~M Dai},
		\bibinfo{person}{Rafal Jozefowicz}, {and} \bibinfo{person}{Samy Bengio}.}
	\bibinfo{year}{2015}\natexlab{}.
	\newblock \showarticletitle{Generating sentences from a continuous space}.
	\newblock \bibinfo{journal}{\emph{arXiv preprint arXiv:1511.06349}}
	(\bibinfo{year}{2015}).
	\newblock
	
	
	\bibitem[\protect\citeauthoryear{Boyd and Vandenberghe}{Boyd and
		Vandenberghe}{2004}]%
	{boyd2004convex}
	\bibfield{author}{\bibinfo{person}{Stephen Boyd} {and} \bibinfo{person}{Lieven
			Vandenberghe}.} \bibinfo{year}{2004}\natexlab{}.
	\newblock \bibinfo{booktitle}{\emph{Convex optimization}}.
	\newblock \bibinfo{publisher}{Cambridge university press}.
	\newblock
	
	
	\bibitem[\protect\citeauthoryear{Brants}{Brants}{2000}]%
	{brants2000tnt}
	\bibfield{author}{\bibinfo{person}{Thorsten Brants}.}
	\bibinfo{year}{2000}\natexlab{}.
	\newblock \showarticletitle{TnT: a statistical part-of-speech tagger}. In
	\bibinfo{booktitle}{\emph{Proceedings of the sixth conference on Applied
			natural language processing}}. ACL, \bibinfo{pages}{224--231}.
	\newblock
	
	
	\bibitem[\protect\citeauthoryear{Chaudhuri, Monteleoni, and Sarwate}{Chaudhuri
		et~al\mbox{.}}{2011}]%
	{chaudhuri2011differentially}
	\bibfield{author}{\bibinfo{person}{Kamalika Chaudhuri}, \bibinfo{person}{Claire
			Monteleoni}, {and} \bibinfo{person}{Anand~D Sarwate}.}
	\bibinfo{year}{2011}\natexlab{}.
	\newblock \showarticletitle{Differentially private empirical risk
		minimization}. In \bibinfo{booktitle}{\emph{JMLR}},
	Vol.~\bibinfo{volume}{12}.
	\newblock
	
	
	\bibitem[\protect\citeauthoryear{Cho, Van~Merri{\"e}nboer, Gulcehre, Bahdanau,
		Bougares, Schwenk, and Bengio}{Cho et~al\mbox{.}}{2014}]%
	{cho2014learning}
	\bibfield{author}{\bibinfo{person}{Kyunghyun Cho}, \bibinfo{person}{Bart
			Van~Merri{\"e}nboer}, \bibinfo{person}{Caglar Gulcehre},
		\bibinfo{person}{Dzmitry Bahdanau}, \bibinfo{person}{Fethi Bougares},
		\bibinfo{person}{Holger Schwenk}, {and} \bibinfo{person}{Yoshua Bengio}.}
	\bibinfo{year}{2014}\natexlab{}.
	\newblock \showarticletitle{Learning phrase representations using RNN
		encoder-decoder for statistical machine translation}.
	\newblock \bibinfo{journal}{\emph{arXiv preprint arXiv:1406.1078}}
	(\bibinfo{year}{2014}).
	\newblock
	
	
	\bibitem[\protect\citeauthoryear{dos Santos and Gatti}{dos Santos and
		Gatti}{2014}]%
	{dos2014deep}
	\bibfield{author}{\bibinfo{person}{Cicero dos Santos} {and}
		\bibinfo{person}{Maira Gatti}.} \bibinfo{year}{2014}\natexlab{}.
	\newblock \showarticletitle{Deep convolutional neural networks for sentiment
		analysis of short texts}. In \bibinfo{booktitle}{\emph{Proceedings of
			Computational Linguistics}}.
	\newblock
	
	
	\bibitem[\protect\citeauthoryear{Dwork}{Dwork}{2008}]%
	{dwork2008differential}
	\bibfield{author}{\bibinfo{person}{Cynthia Dwork}.}
	\bibinfo{year}{2008}\natexlab{}.
	\newblock \showarticletitle{Differential privacy: A survey of results}. In
	\bibinfo{booktitle}{\emph{International Conference on Theory and Applications
			of Models of Computation}}. Springer, \bibinfo{pages}{1--19}.
	\newblock
	
	
	\bibitem[\protect\citeauthoryear{Dwork, Roth, et~al\mbox{.}}{Dwork
		et~al\mbox{.}}{2014}]%
	{dwork2014algorithmic}
	\bibfield{author}{\bibinfo{person}{Cynthia Dwork}, \bibinfo{person}{Aaron
			Roth}, {et~al\mbox{.}}} \bibinfo{year}{2014}\natexlab{}.
	\newblock \showarticletitle{The algorithmic foundations of differential
		privacy}.
	\newblock \bibinfo{journal}{\emph{Foundations and Trends in Theoretical
			Computer Science}} (\bibinfo{year}{2014}).
	\newblock
	
	
	\bibitem[\protect\citeauthoryear{Fung, Wang, Chen, and Philip}{Fung
		et~al\mbox{.}}{2010}]%
	{fung2010privacy}
	\bibfield{author}{\bibinfo{person}{Benjamin~CM Fung}, \bibinfo{person}{K Wang},
		\bibinfo{person}{R Chen}, {and} \bibinfo{person}{S~Yu Philip}.}
	\bibinfo{year}{2010}\natexlab{}.
	\newblock \showarticletitle{Privacy-preserving data publishing: A survey of
		recent developments}.
	\newblock \bibinfo{journal}{\emph{Comput. Surveys}} \bibinfo{volume}{42},
	\bibinfo{number}{4} (\bibinfo{year}{2010}).
	\newblock
	
	
	\bibitem[\protect\citeauthoryear{Gervais, Shokri, Singla, Capkun, and
		Lenders}{Gervais et~al\mbox{.}}{2014}]%
	{gervais2014quantifying}
	\bibfield{author}{\bibinfo{person}{Arthur Gervais}, \bibinfo{person}{Reza
			Shokri}, \bibinfo{person}{Adish Singla}, \bibinfo{person}{Srdjan Capkun},
		{and} \bibinfo{person}{Vincent Lenders}.} \bibinfo{year}{2014}\natexlab{}.
	\newblock \showarticletitle{Quantifying web-search privacy}. In
	\bibinfo{booktitle}{\emph{Proceedings of ACM SIGSAC on CCS}}.
	\newblock
	
	
	\bibitem[\protect\citeauthoryear{Goodfellow, Pouget-Abadie, Mirza, Xu,
		Warde-Farley, Ozair, Courville, and Bengio}{Goodfellow et~al\mbox{.}}{2014}]%
	{goodfellow2014generative}
	\bibfield{author}{\bibinfo{person}{Ian Goodfellow}, \bibinfo{person}{Jean
			Pouget-Abadie}, \bibinfo{person}{Mehdi Mirza}, \bibinfo{person}{Bing Xu},
		\bibinfo{person}{David Warde-Farley}, \bibinfo{person}{Sherjil Ozair},
		\bibinfo{person}{Aaron Courville}, {and} \bibinfo{person}{Yoshua Bengio}.}
	\bibinfo{year}{2014}\natexlab{}.
	\newblock \showarticletitle{Generative adversarial nets}. In
	\bibinfo{booktitle}{\emph{Advances in neural information processing
			systems}}. \bibinfo{pages}{2672--2680}.
	\newblock
	
	
	\bibitem[\protect\citeauthoryear{Gotz, Machanavajjhala, Wang, Xiao, and
		Gehrke}{Gotz et~al\mbox{.}}{2012}]%
	{gotz2012publishing}
	\bibfield{author}{\bibinfo{person}{Michaela Gotz}, \bibinfo{person}{Ashwin
			Machanavajjhala}, \bibinfo{person}{Guozhang Wang}, \bibinfo{person}{Xiaokui
			Xiao}, {and} \bibinfo{person}{Johannes Gehrke}.}
	\bibinfo{year}{2012}\natexlab{}.
	\newblock \showarticletitle{Publishing search logs a comparative study of
		privacy guarantees}.
	\newblock \bibinfo{journal}{\emph{IEEE Transactions on Knowledge and Data
			Engineering}} \bibinfo{volume}{24}, \bibinfo{number}{3}
	(\bibinfo{year}{2012}).
	\newblock
	
	
	\bibitem[\protect\citeauthoryear{Hakkini-Tur, Tur, et~al\mbox{.}}{Hakkini-Tur
		et~al\mbox{.}}{2006}]%
	{hakkini2006sanitization}
	\bibfield{author}{\bibinfo{person}{Dilek Hakkini-Tur}, \bibinfo{person}{Gˆkhan
			Tur}, {et~al\mbox{.}}} \bibinfo{year}{2006}\natexlab{}.
	\newblock \showarticletitle{Sanitization and anonymization of document
		repositories}.
	\newblock In \bibinfo{booktitle}{\emph{Web and information security}}.
	\bibinfo{publisher}{IGI Global}, \bibinfo{pages}{133--148}.
	\newblock
	
	
	\bibitem[\protect\citeauthoryear{Hitaj, Ateniese, and Perez-Cruz}{Hitaj
		et~al\mbox{.}}{2017}]%
	{hitaj2017deep}
	\bibfield{author}{\bibinfo{person}{Briland Hitaj}, \bibinfo{person}{Giuseppe
			Ateniese}, {and} \bibinfo{person}{Fernando Perez-Cruz}.}
	\bibinfo{year}{2017}\natexlab{}.
	\newblock \showarticletitle{Deep models under the GAN: information leakage from
		collaborative deep learning}. In \bibinfo{booktitle}{\emph{Proceedings of ACM
			SIGSAC Conference on Computer and Communications Security}}.
	\newblock
	
	
	\bibitem[\protect\citeauthoryear{Hovy, Johannsen, and S{\o}gaard}{Hovy
		et~al\mbox{.}}{2015}]%
	{hovy2015user}
	\bibfield{author}{\bibinfo{person}{Dirk Hovy}, \bibinfo{person}{Anders
			Johannsen}, {and} \bibinfo{person}{Anders S{\o}gaard}.}
	\bibinfo{year}{2015}\natexlab{}.
	\newblock \showarticletitle{User review sites as a resource for large-scale
		sociolinguistic studies}. In \bibinfo{booktitle}{\emph{Proceedings of WWW}}.
	\newblock
	
	
	\bibitem[\protect\citeauthoryear{Hovy and S{\o}gaard}{Hovy and
		S{\o}gaard}{2015}]%
	{hovy2015tagging}
	\bibfield{author}{\bibinfo{person}{Dirk Hovy} {and} \bibinfo{person}{Anders
			S{\o}gaard}.} \bibinfo{year}{2015}\natexlab{}.
	\newblock \showarticletitle{Tagging performance correlates with author age}. In
	\bibinfo{booktitle}{\emph{Proceedings of ACL}}.
	\newblock
	
	
	\bibitem[\protect\citeauthoryear{Howe and Nissenbaum}{Howe and
		Nissenbaum}{2009}]%
	{howe2009trackmenot}
	\bibfield{author}{\bibinfo{person}{Daniel~C Howe} {and} \bibinfo{person}{Helen
			Nissenbaum}.} \bibinfo{year}{2009}\natexlab{}.
	\newblock \showarticletitle{TrackMeNot: Resisting surveillance in web search}.
	\newblock \bibinfo{journal}{\emph{Lessons from the Identity trail: Anonymity,
			privacy, and identity in a networked society}}  \bibinfo{volume}{23}
	(\bibinfo{year}{2009}), \bibinfo{pages}{417--436}.
	\newblock
	
	
	\bibitem[\protect\citeauthoryear{J{\o}rgensen, Hovy, and
		S{\o}gaard}{J{\o}rgensen et~al\mbox{.}}{2016}]%
	{jorgensen2016learning}
	\bibfield{author}{\bibinfo{person}{Anna J{\o}rgensen}, \bibinfo{person}{Dirk
			Hovy}, {and} \bibinfo{person}{Anders S{\o}gaard}.}
	\bibinfo{year}{2016}\natexlab{}.
	\newblock \showarticletitle{Learning a POS tagger for AAVE-like language}. In
	\bibinfo{booktitle}{\emph{Proceedings of ACL: Human Language Technologies}}.
	\newblock
	
	
	\bibitem[\protect\citeauthoryear{Kifer and Machanavajjhala}{Kifer and
		Machanavajjhala}{2011}]%
	{kifer2011no}
	\bibfield{author}{\bibinfo{person}{Daniel Kifer} {and} \bibinfo{person}{Ashwin
			Machanavajjhala}.} \bibinfo{year}{2011}\natexlab{}.
	\newblock \showarticletitle{No free lunch in data privacy}. In
	\bibinfo{booktitle}{\emph{Proceedings of ACM SIGMOD International Conference
			on Management of data}}.
	\newblock
	
	
	\bibitem[\protect\citeauthoryear{Kingma and Welling}{Kingma and
		Welling}{2013}]%
	{kingma2013auto}
	\bibfield{author}{\bibinfo{person}{Diederik~P Kingma} {and}
		\bibinfo{person}{Max Welling}.} \bibinfo{year}{2013}\natexlab{}.
	\newblock \showarticletitle{Auto-encoding variational bayes}.
	\newblock \bibinfo{journal}{\emph{arXiv preprint arXiv:1312.6114}}
	(\bibinfo{year}{2013}).
	\newblock
	
	
	\bibitem[\protect\citeauthoryear{Korolova, Kenthapadi, Mishra, and
		Ntoulas}{Korolova et~al\mbox{.}}{2009}]%
	{korolova2009releasing}
	\bibfield{author}{\bibinfo{person}{Aleksandra Korolova},
		\bibinfo{person}{Krishnaram Kenthapadi}, \bibinfo{person}{Nina Mishra}, {and}
		\bibinfo{person}{Alexandros Ntoulas}.} \bibinfo{year}{2009}\natexlab{}.
	\newblock \showarticletitle{Releasing search queries and clicks privately}. In
	\bibinfo{booktitle}{\emph{WWW}}.
	\newblock
	
	
	\bibitem[\protect\citeauthoryear{Le and Mikolov}{Le and Mikolov}{2014}]%
	{le2014distributed}
	\bibfield{author}{\bibinfo{person}{Quoc Le} {and} \bibinfo{person}{Tomas
			Mikolov}.} \bibinfo{year}{2014}\natexlab{}.
	\newblock \showarticletitle{Distributed representations of sentences and
		documents}. In \bibinfo{booktitle}{\emph{International Conference on Machine
			Learning}}. \bibinfo{pages}{1188--1196}.
	\newblock
	
	
	\bibitem[\protect\citeauthoryear{Li, Baldwin, and Cohn}{Li
		et~al\mbox{.}}{2018}]%
	{li2018towards}
	\bibfield{author}{\bibinfo{person}{Yitong Li}, \bibinfo{person}{Timothy
			Baldwin}, {and} \bibinfo{person}{Trevor Cohn}.}
	\bibinfo{year}{2018}\natexlab{}.
	\newblock \showarticletitle{Towards Robust and Privacy-preserving Text
		Representations}.
	\newblock  (\bibinfo{year}{2018}).
	\newblock
	
	
	\bibitem[\protect\citeauthoryear{Lui and Baldwin}{Lui and Baldwin}{2012}]%
	{lui2012langid}
	\bibfield{author}{\bibinfo{person}{Marco Lui} {and} \bibinfo{person}{Timothy
			Baldwin}.} \bibinfo{year}{2012}\natexlab{}.
	\newblock \showarticletitle{langid. py: An off-the-shelf language
		identification tool}. In \bibinfo{booktitle}{\emph{Proceedings of the ACL
			2012 system demonstrations}}.
	\newblock
	
	
	\bibitem[\protect\citeauthoryear{McSherry and Mironov}{McSherry and
		Mironov}{2009}]%
	{mcsherry2009differentially}
	\bibfield{author}{\bibinfo{person}{Frank McSherry} {and} \bibinfo{person}{Ilya
			Mironov}.} \bibinfo{year}{2009}\natexlab{}.
	\newblock \showarticletitle{Differentially private recommender systems:
		building privacy into the net}. In \bibinfo{booktitle}{\emph{Proceedings of
			the 15th ACM SIGKDD}}.
	\newblock
	
	
	\bibitem[\protect\citeauthoryear{Meng, Wang, Shu, Li, Chen, Liu, and
		Zhang}{Meng et~al\mbox{.}}{2018}]%
	{meng2018personalized}
	\bibfield{author}{\bibinfo{person}{Xuying Meng}, \bibinfo{person}{Suhang Wang},
		\bibinfo{person}{Kai Shu}, \bibinfo{person}{Jundong Li}, \bibinfo{person}{Bo
			Chen}, \bibinfo{person}{Huan Liu}, {and} \bibinfo{person}{Yujun Zhang}.}
	\bibinfo{year}{2018}\natexlab{}.
	\newblock \showarticletitle{Personalized privacy-preserving social
		recommendation}. In \bibinfo{booktitle}{\emph{Proceedings of Thirty-Second
			AAAI Conference on Artificial Intelligence.}}
	\newblock
	
	
	\bibitem[\protect\citeauthoryear{Mukherjee and Liu}{Mukherjee and Liu}{2010}]%
	{mukherjee2010improving}
	\bibfield{author}{\bibinfo{person}{Arjun Mukherjee} {and} \bibinfo{person}{Bing
			Liu}.} \bibinfo{year}{2010}\natexlab{}.
	\newblock \showarticletitle{Improving gender classification of blog authors}.
	In \bibinfo{booktitle}{\emph{Proceedings of the 2010 conference on ACL
			EMNLP}}.
	\newblock
	
	
	\bibitem[\protect\citeauthoryear{Narayanan and Shmatikov}{Narayanan and
		Shmatikov}{2008}]%
	{narayanan2008robust}
	\bibfield{author}{\bibinfo{person}{Arvind Narayanan} {and}
		\bibinfo{person}{Vitaly Shmatikov}.} \bibinfo{year}{2008}\natexlab{}.
	\newblock \showarticletitle{Robust de-anonymization of large sparse datasets}.
	In \bibinfo{booktitle}{\emph{IEEE Symposium on Security and Privacy}}.
	\newblock
	
	
	\bibitem[\protect\citeauthoryear{Petrov, Das, and McDonald}{Petrov
		et~al\mbox{.}}{2012}]%
	{petrov2012universal}
	\bibfield{author}{\bibinfo{person}{Slav Petrov}, \bibinfo{person}{Dipanjan
			Das}, {and} \bibinfo{person}{Ryan McDonald}.}
	\bibinfo{year}{2012}\natexlab{}.
	\newblock \showarticletitle{A Universal Part-of-Speech Tagset}. In
	\bibinfo{booktitle}{\emph{Proceedings of Language Resources and Evaluation
			(LREC)}}.
	\newblock
	
	
	\bibitem[\protect\citeauthoryear{Potthast, Rangel, Tschuggnall, Stamatatos,
		Rosso, and Stein}{Potthast et~al\mbox{.}}{2017}]%
	{potthast2017overview}
	\bibfield{author}{\bibinfo{person}{Martin Potthast}, \bibinfo{person}{Francisco
			Rangel}, \bibinfo{person}{Michael Tschuggnall}, \bibinfo{person}{Efstathios
			Stamatatos}, \bibinfo{person}{Paolo Rosso}, {and} \bibinfo{person}{Benno
			Stein}.} \bibinfo{year}{2017}\natexlab{}.
	\newblock \showarticletitle{Overview of PAN'17}. In
	\bibinfo{booktitle}{\emph{International Conference of the Cross-Language
			Evaluation Forum for European Languages}}.
	\newblock
	
	
	\bibitem[\protect\citeauthoryear{Shang, Lu, and Li}{Shang
		et~al\mbox{.}}{2015}]%
	{shang2015neural}
	\bibfield{author}{\bibinfo{person}{Lifeng Shang}, \bibinfo{person}{Zhengdong
			Lu}, {and} \bibinfo{person}{Hang Li}.} \bibinfo{year}{2015}\natexlab{}.
	\newblock \showarticletitle{Neural responding machine for short-text
		conversation}.
	\newblock \bibinfo{journal}{\emph{arXiv preprint arXiv:1503.02364}}
	(\bibinfo{year}{2015}).
	\newblock
	
	
	\bibitem[\protect\citeauthoryear{Volkova, Bachrach, Armstrong, and
		Sharma}{Volkova et~al\mbox{.}}{2015}]%
	{volkova2015inferring}
	\bibfield{author}{\bibinfo{person}{Svitlana Volkova}, \bibinfo{person}{Yoram
			Bachrach}, \bibinfo{person}{Michael Armstrong}, {and} \bibinfo{person}{Vijay
			Sharma}.} \bibinfo{year}{2015}\natexlab{}.
	\newblock \showarticletitle{Inferring Latent User Properties from Texts
		Published in Social Media.}. In \bibinfo{booktitle}{\emph{Proceedings of
			Twenty-Ninth AAAI Conference on Artificial Intelligence.}}
	\newblock
	
	
	\bibitem[\protect\citeauthoryear{Xiao, Chen, and Tan}{Xiao
		et~al\mbox{.}}{2014}]%
	{xiao2014differentially}
	\bibfield{author}{\bibinfo{person}{Qian Xiao}, \bibinfo{person}{Rui Chen},
		{and} \bibinfo{person}{Kian-Lee Tan}.} \bibinfo{year}{2014}\natexlab{}.
	\newblock \showarticletitle{Differentially private network data release via
		structural inference}. In \bibinfo{booktitle}{\emph{Proceedings of the 20th
			ACM SIGKDD}}.
	\newblock
	
	
	\bibitem[\protect\citeauthoryear{Zhang, Sun, Zhang, and Zhang}{Zhang
		et~al\mbox{.}}{2018}]%
	{TextAnonymization}
	\bibfield{author}{\bibinfo{person}{Jinxue Zhang}, \bibinfo{person}{Jingchao
			Sun}, \bibinfo{person}{Rui Zhang}, {and} \bibinfo{person}{Yanchao Zhang}.}
	\bibinfo{year}{2018}\natexlab{}.
	\newblock \showarticletitle{Privacy-Preserving Social Media Data Outsourcing}.
	In \bibinfo{booktitle}{\emph{Proceedings of IEEE INFOCOM}}.
	\newblock
	
	
	\bibitem[\protect\citeauthoryear{Zhang, Yang, and Singh}{Zhang
		et~al\mbox{.}}{2016}]%
	{zhang2016anonymizing}
	\bibfield{author}{\bibinfo{person}{Sicong Zhang}, \bibinfo{person}{Hui Yang},
		{and} \bibinfo{person}{Lisa Singh}.} \bibinfo{year}{2016}\natexlab{}.
	\newblock \showarticletitle{Anonymizing query logs by differential privacy}. In
	\bibinfo{booktitle}{\emph{Proceedings of ACM SIGIR}}.
	\newblock
	
	
\end{thebibliography}

%%% -*-BibTeX-*-
%%% Do NOT edit. File created by BibTeX with style
%%% ACM-Reference-Format-Journals [18-Jan-2012].

\ifx \showCODEN    \undefined \def \showCODEN     #1{\unskip}     \fi
\ifx \showDOI      \undefined \def \showDOI       #1{#1}\fi
\ifx \showISBNx    \undefined \def \showISBNx     #1{\unskip}     \fi
\ifx \showISBNxiii \undefined \def \showISBNxiii  #1{\unskip}     \fi
\ifx \showISSN     \undefined \def \showISSN      #1{\unskip}     \fi
\ifx \showLCCN     \undefined \def \showLCCN      #1{\unskip}     \fi
\ifx \shownote     \undefined \def \shownote      #1{#1}          \fi
\ifx \showarticletitle \undefined \def \showarticletitle #1{#1}   \fi
\ifx \showURL      \undefined \def \showURL       {\relax}        \fi
% The following commands are used for tagged output and should be
% invisible to TeX

\end{document}